\documentclass[12pt,preprint]{aastex}

\shorttitle{The Upper Atmosphere of HD17156b}
\shortauthors{Koskinen et al.}

\begin{document}

\title{The Upper Atmosphere of HD17156b}

\author{T.T.Koskinen, A.D.Aylward, S.Miller}
\affil{Department of Physics and Astronomy, University College London, \\
       Gower Street, London WC1E 6BT, UK}
\email{tommi@apl.ucl.ac.uk}

\begin{abstract}

HD17156b is a newly-found transiting extrasolar giant planet (EGP) that orbits its G-type 
host star in a highly eccentric orbit (e~$\sim$0.67) with an orbital semi-major axis 
of 0.16 AU.  Its period, 21.2 Earth days, is the longest among the known transiting planets.  
The atmosphere of the planet undergoes a 27-fold variation in stellar irradiation during each 
orbit, making it an interesting subject for atmospheric modelling.  
We have used a three-dimensional model of the upper atmosphere and ionosphere for 
extrasolar gas giants in order to simulate the progress of HD17156b along its eccentric 
orbit.  Here we present the results of these simulations and discuss the stability, 
circulation, and composition in its upper atmosphere.  Contrary to the 
well-known transiting planet HD209458b, we find that the atmosphere of HD17156b is 
unlikely to escape hydrodynamically at any point along the orbit, even if the upper 
atmosphere is almost entirely composed of atomic hydrogen and H$^+$, and infrared cooling 
by H$_{3}^{+}$ ions is negligible.  The nature of the upper atmosphere is sensitive to 
to the composition of the thermosphere, and in particular to the mixing ratio of H$_2$, 
as the availability of H$_2$ regulates radiative cooling.  In light of different 
simulations we make specific predictions about the thermosphere-ionosphere system of 
HD17156b that can potentially be verified by observations.  
    
\end{abstract}

\keywords{hydrodynamics --- planetary systems --- instabilities --- plasmas --- infrared: general}

\section{Introduction}

HD17156b is a transiting, extrasolar giant planet (EGP) that was discovered recently by \citet{fisch07} 
as a part of the N2K program, which is a survey of metal-rich stars intended to 
identify short-period planets \citep{fisch05}.  The transit was first detected by a group of 
amateur astronomers \citep{barbieri07}, belonging to the \textit{Transitsearch.org} network, which conducts 
photometric follow-up observations of known exoplanets that have been detected with the radial 
velocity method \citep{seagroves03}.  Following on these initial detections, improved orbital and 
planetary parameters for HD17156b have recently been released by two groups, \citet{gillon07} and 
\citet{irwin08}, both of whom performed independent transit photometry on the system with different 
ground-based telescopes.  For this work, we have mostly adopted the parameters given in \citet{gillon07}.   

HD17156b orbits a G0 star at a distance of $\sim$78 pc from Earth.  Based on the observations 
of \citet{fisch07}, the bolometric luminosity of the star is 2.6~L$_{Sun}$ and its age is 
$\sim$5.7 Gyr.  Its effective temperature, T$_{eff} =$~6079~K, metallicity $\lbrack$Fe/H$\rbrack$~=~0.24, 
radius R$_* =$~1.47 R$_{Sun}$, and mass, M$_* =$~1.2 M$_{Sun}$.  \citet{fisch07} note that 
the absence of Ca II H \& K line emission in the spectrum of the star indicates low chromospheric activity.  
They measured S$_{HK} =$~0.15 and R'$_{HK} =$~-5.04, and derived a rotational period P$_{ROT} =$~12.8 days 
for the star.  
These observations have implications on the likely magnitude of the stellar XUV emissions that 
are an important energy source in the upper atmosphere of the planet.      

HD17156b is in many ways a remarkable planet.  The semi-major axis of its orbit is a~$=$~0.16 AU, 
and the orbit is highly eccentric, with an eccentricity of e $=$ 0.67.  The orbital period 
of P $=$ 21.2 days is the longest among the known transiting planets.  Due to these 
characteristics, the planet faces a 27-fold stellar flux variation during each orbit, as it moves 
from $\sim$0.27 AU at apastron to 0.052 AU at periastron.  In addition to its curious orbit, HD17156b 
is unique in other ways as well.  It is over three times as massive as Jupiter, but its radius is comparable 
to that of Jupiter.  The average density of the planet is thus $\rho_p =$ 3.47 $\rho_{Jup}$, and this 
makes it one of the densest gas giants known at present.

Due to the high eccentricity of its orbit, and the relatively long semi-major axis, HD17156b 
is not likely to be rotationally synchronised to its host star.  However, the strong tidal forces 
between the planet and the star during periastron are likely to have driven it into 
`pseudo-synchronous' rotation \citep{barbieri07}.  `Pseudo-synchronous' rotation implies 
synchronisation during periastron, but asynchronous rotation elsewhere along the orbit.  
The character of the orbit and the rotation of the planet around its axis make HD17156b an 
interesting subject for dynamical studies of its atmosphere.   
\citet{irwin08} simulated the meteorology in the photosphere of the planet 
by using the climate model of \citet{langton08}, and used the results 
to estimate the phase-dependent 8 $\mu$m infrared emissions from the planet.  
In this paper, we present three-dimensional simulations of the upper atmosphere and 
ionosphere of the planet, and investigate the stability of the atmosphere against thermal 
evaporation. 

Absorption by atomic hydrogen in the atmosphere of another close-in EGP, 
HD209458b, observed in the stellar Lyman~$\alpha$ emission line during the transit of the planet 
across its host star, indicates that it is 
surrounded by an extended atmosphere, composed mainly of atomic hydrogen and H$^+$ 
\citep{vidalmadjar03,vidalmadjar08}.  The observations indicate that the atmosphere is 
escaping hydrodynamically with a minimum mass loss rate of 10$^{10}$ gs$^{-1}$.  
This conclusion is also supported by the tentative detection of oxygen and ionised carbon in 
the upper atmosphere of the planet \citep{vidalmadjar04}, the detection of hot atomic hydrogen 
in the thermosphere \citep{ballester07} and numerous modelling studies 
\citep[eg.][]{yelle04,yelle06,tian05,garciamunoz07,erkaev07}.  A recent attempt to constrain the 
mass loss from the planet further yielded a rate of (1.1 $\pm$ 0.3) $\times$10$^{10}$ gs$^{-1}$ 
\citep{schneiter07}.  This estimate is based on modelling the Lyman~$\alpha$ absorption, taking 
into account the interaction of the exosphere with the incoming stellar wind.

While it is reasonable to suggest that the atmospheres of other close-in EGPs should undergo 
hydrodynamic escape as well, compelling observational evidence only exists for HD209458b.  
In general, the stability of the atmosphere against hydrodynamic escape depends on the conditions in 
the upper atmosphere and ionosphere, which are heated mainly by the absorption of stellar 
XUV radiation.  Models of the thermosphere can thus be used to assess the overall stability 
of EGP atmospheres.  Using a three-dimensional thermosphere-ionosphere model for extrasolar 
giant planets (EXOTIM), we recently published simulations of the upper atmospheres of 
Jupiter-type EGPs orbiting a solar-type host star at different distances \citep{koskinen07a}.  
These simulations 
indicate that infrared cooling from H$_{3}^{+}$ ions may play an important role in cooling the 
thermospheres of EGPs.  This H$_{3}^{+}$ cooling effect was modelled earlier by \citep{yelle04, 
yelle06} for HD209458b in a context of a one-dimensional non-hydrostatic model.  Our calculations 
extended this effect to three dimensions and a wider sample of orbital distances.  
We also identified a sharp stability limit for the atmosphere 
of a Jupiter-type exoplanet orbiting the Sun at $\sim$0.15 AU \citep{koskinen07b}.  Within 
this limit molecular hydrogen breaks down in the upper atmosphere due to thermal 
dissociation, the atmosphere expands, and begins to escape hydrodynamically.  Outside the 
limit, the thermosphere is effectively cooled by H$_{3}^{+}$, and the atmosphere does not  
undergo fast hydrodynamic escape.

In this context, HD17156b is an interesting subject for similar modelling studies.  During 
periastron, the atmosphere of the planet should absorb enough XUV radiation from the star to undergo 
hydrodynamic escape, whereas at apastron distances we would not expect this to be the case. 
Of course, this kind of reasoning is oversimplistic.  The planet may not 
spend long enough near periastron for the atmosphere to heat up, or it may not have 
enough time to cool down during apastron.  As a result, the atmosphere may be stable 
throughout the whole orbit, or it may escape hydrodynamically at every point.  Also, 
HD17156b is much heavier than Jupiter (that we used as a template in our earlier 
work).  Consequently, the scale heights in the atmosphere are shorter, and due to higher 
gravity, particles need more thermal kinetic energy to escape the atmosphere.  Higher gravity 
may thus help to maintain stability even at close-in distances during periastron.       

In order to make predictions about the upper atmosphere, ionosphere, and 
potential infrared emissions, we have modified EXOTIM to simulate the progress of HD17156b 
along its eccentric orbit.  The details of the new model are discussed in 
Section~\ref{sc:model}.  Section~\ref{sc:results} describes the results of the 
modelling along with some caveats, and Section~\ref{sc:discussion} discusses the implications 
of our results in the context of potential observations and relates them to earlier work 
on this subject. 

\section{The Model}
\label{sc:model}

\subsection{Basic equations}

The model solves the three-dimensional Navier-Stokes equations of continuity, momentum and 
energy by explicit time-integration, assuming that the thermosphere is dominated by H$_2$, 
He and H and that it is in hydrostatic equilibrium throughout.  The basic equations are solved 
on a non-inertial Eulerian, corotating spherical grid, using spherical pressure coordinates.  
The details of the equations, and the model in general, are described extensively in 
\citet{koskinen07a} and are thus not repeated here. 
However, since the publication of that paper, we have added a number 
of improvements and modifications to the model, which are introduced below.  Some of these 
changes were also described in the Supplementary Information of \citet{koskinen07b}.  It should 
be noted that the resulting equations resemble the primitive equations of meteorology, with 
the exception that they include terms due to molecular diffusion, which are important in the 
rare gas of the upper atmosphere.  Also, our equations fully include the curvature terms arising 
from spherical coordinates.

As before, we solve for the horizontal winds from the momentum equations, accounting for 
advection, geopotential gradients, the Coriolis force and viscosity due to molecular 
diffusion.  Once the horizontal velocities are known, we solve for the vertical velocities 
from the continuity equation.  The temperatures are solved from the energy equation, 
which accounts for the advection of energy around the atmosphere by circulation, 
the absorption of stellar X-rays and EUV (XUV) radiation, infrared cooling (by H$_{3}^{+}$), heating and 
cooling due to adiabatic contraction and expansion, vertical and horizontal conduction of heat 
and viscous heating.  The basic equations of motion are coupled to continuity equations 
for neutral mass fractions that account for vertical and horizontal advection, vertical diffusion and 
neutral chemistry.  The numerical solution is based on finite-difference approximations 
of spatial derivatives and forward-time integration.  This scheme is prone to numerical 
instabilities, which we filter out by using the two-step smoothing element described by 
\citet{shapiro70}.  This element has been adapted for spherical geometry and vector equations, 
where appropriate.  We only apply smoothing in the horizontal direction, and use the filter 
every 144 s of runtime on temperatures and winds, and every 36 s of runtime on mass mixing 
ratios of the neutral species.  Generally, the timestep in the simulations varies between 1-2 s.  

\subsection{Composition}

As an improvement on the previous model, we have updated the photochemistry and 
calculated new values for the absorption cross sections of H$_2$, He and H throughout the 
XUV spectrum.  The new reaction list with appropriate references is shown in 
Table~\ref{table:reactions}.  Instead of using coefficients of heat conduction and 
molecular diffusion for atomic hydrogen given by \citet{achilleos98}, we have adopted 
values from \citet{banks73}.  The coefficients for other species are as before.  We note that 
the changes do not affect the results presented in \citet{koskinen07a} significantly.

We determine ion densities by calculating photoionisation by stellar XUV radiation, 
and model the subsequent photochemistry by using the reactions listed in Table~\ref{table:reactions} 
and assuming photochemical equilibrium.  In doing so, we effectively ignore ion transport, 
thus assuming that ion lifetimes are short.  Also, we calculate the overall neutral density 
from the ideal gas law, assuming that photoionisation has negligible \textit{direct} 
impact on neutral mass fractions. This assumption is valid as long as ion densities are much 
smaller than neutral densities.

In general, the number density of ion species $i$, $n_i$ is given by:

\begin{equation}
\frac{dn_i}{dt} = P_i + L_i n_i
\label{eq:ion_density}
\end{equation}
where $P_i$ and $L_i$ are the production and loss rates of species $i$, respectively.  
For a tidally locked planet in photochemical equilibrium, we have $P_i + L_i n_i = 0$, 
and the resulting series of non-linear, simultaneous equations can be solved 
by using Newton-Raphson iteration \citep{koskinen07a, press92}.  
However, if the planet rotates asynchronously, the rate of change in density $n_i$ is 
not zero in the Eulerian, corotating frame.  For such planets, our previous model used 
a crude time-integration technique:

\begin{equation}
n_i (t + \delta t) = n(t) + (P_i + L_i n_i) \delta t
\end{equation}
This method is unstable for short ion lifetimes, which are typical for H$_{2}^{+}$, 
H$_{3}^{+}$, and He$^+$.  In general, it requires unrealistically short time steps, and 
it also forced us to assume that all H$_{2}^{+}$ is immediately turned into 
H$_{3}^{+}$.  Thus we were unable to model the H$_{2}^{+}$ densities self-consistently 
for asynchronously rotating planets.  Clearly, such a limitation would have been 
unacceptable for planet in `pseudo-synchronous' rotation such as HD17156b. 

Equation (\ref{eq:ion_density}) can be integrated with respect to time 
by using the method of integrating factors, and thus we used an improved time-integration 
formula:

\begin{equation}
n_i (t+\delta t) = \frac{P_i}{L_i} (1-e^{-L_i \delta t}) + n_i(t) e^{-L_i \delta t}
\label{eq:ion_result}
\end{equation}
As long as the time step $\delta t$ is kept reasonably short, this method produces 
accurate ion densities for asynhronously rotating atmospheres in photochemical 
equilibrium.  It also allows us to fully include H$_{2}^{+}$ in the calculations.   

\subsection{XUV Heating}

Previously the model included stellar radiation between 5.0 and 105 nm.  We have 
extended this range to cover X-rays, and now include radiation from 0.1 to 105 nm.  
We note that for XUV emissions typical of the current Sun, the inclusion of X-rays 
in our input has almost no impact on the properties of the thermosphere compared 
to models that do not include X-rays.  Some authors have proposed that X-rays have 
a significant impact on the atmospheres of EGPs orbiting young G stars 
\citep[eg.][]{cecchi06,penz08}, and this may well be the case because a more 
significant fraction of the total XUV flux from young stars is in X-rays than from 
mature solar-type stars \citep{ribas05}.  However, it is not true for EGPs orbiting 
mature main sequence G stars such as the Sun. 

We assume that the XUV flux from HD17156b is roughly similar to solar flux, and 
use solar maximum fluxes from the SOLAR2000 model \citep{tobiska00} 
in these simulations.  This assumption is not likely to be accurate, and considerably 
variation in the XUV flux is possible.  Unfortunately, there are no observations or 
models of the XUV emissions from HD17156.  The intensity of the emissions is related to 
the strength of the stellar magnetic dynamo, which depends on the rotation rate 
of the star \citep{ribas05}.  The period of rotation for HD17156 is half that of the Sun, 
so the XUV flux could be slightly higher, but this is speculative.
In the absence of better estimates, we have little choice but to adopt solar emissions 
for the time being.  
Despite this uncertainty, the models will still help us to develop an understanding of 
the relative dynamics and structure in the thermosphere, and perhaps the predictions 
can be tested with future observations.  

It should be noted here that we did not perform detailed energetic calculations 
to work out the heating rate due to the absorption of stellar XUV energy.  Instead, 
we assumed that 50 \% of the absorbed energy is thermalised in a H$_2$-dominated 
atmosphere.  This is consistent with calculations performed by \citet{waite83} for 
Jupiter's upper atmosphere that account for all other energetic processes apart 
from H$_{3}^{+}$ cooling.  Following \citet{yelle04}, we assumed a 10 \% heating 
efficiency in those layers of the thermosphere that are dominated by atomic hydrogen. 
  
\subsection{Infrared Cooling}

The only infrared cooling source included in our model is due to H$_{3}^{+}$ emissions. 
We assume that the upper atmosphere is optically thin in the infrared and 
ignore radiative transfer.  Thus any emitted infrared radiation escapes directly 
to space or to the lower atmosphere, and in both cases contributes to the cooling 
of the thermosphere.  We neglect the possible presence of hydrocarbons or carbon 
monoxide near the lower boundary of the model that may in reality complicate this 
picture.

In order to calculate the total emission rate in local thermodynamic equilibrium 
(LTE), we employ the complete H$_{3}^{+}$ line list published by \citet{neale96}.  
However, the frequency of intermolecular collisions in the upper atmospheres of 
gas giants is generally too low to maintain LTE conditions.  Previously, we 
estimated non-LTE emission rates by multiplying the LTE rates by a 
pressure-dependent correction factor, which was derived experimentally by fitting 
to \textit{Galileo} observations of Jupiter's thermosphere \citep{koskinen07a}.  
As an improvement, we now base the correction factor on detailed balance calculations. 

It would be computationally impossible to perform detailed balance calculations for 
over three million line transitions included in the H$_{3}^{+}$ line list, at every grid 
point in the model, and during each time step.  Instead, we have singled out prominent 
vibrational transitions included in an earlier line list of \citet{dinelli92} that 
account for most of the LTE infrared emissions.  We determined both the LTE and non-LTE 
emission rates for these transitions, adopting detailed balance calculations for the 
non-LTE emissions, and calculated the ratio of the total non-LTE rates to LTE rates 
for a simple grid of temperatures and densities that encompasses the likely values 
within the model.  We then created a table of non-LTE/LTE ratios for different 
temperatures and densities. As the model runs, it interpolates from this table bilinearly 
in order to work out the non-LTE/LTE ratio for each grid point.  
The final emission rate is worked out by multiplying the total LTE emission rate, 
evaluated by using a fourth-order fitted polynomial to the values given by \citet{neale96}, 
by this ratio.  This method essentially assumes that the non-LTE correction is the same 
for the complete line list as it is for the reduced line list of \citet{dinelli92}.  
The resulting ratios are listed for the substellar point of a typical model run, together 
with temperatures and pressures, in Table~\ref{table:nonlte}.  For comparison, 
experimental correction factors given by equation (15) of \citet{koskinen07a} are 
also shown.         

\subsection{The Orbit}

\begin{figure}
\plotone{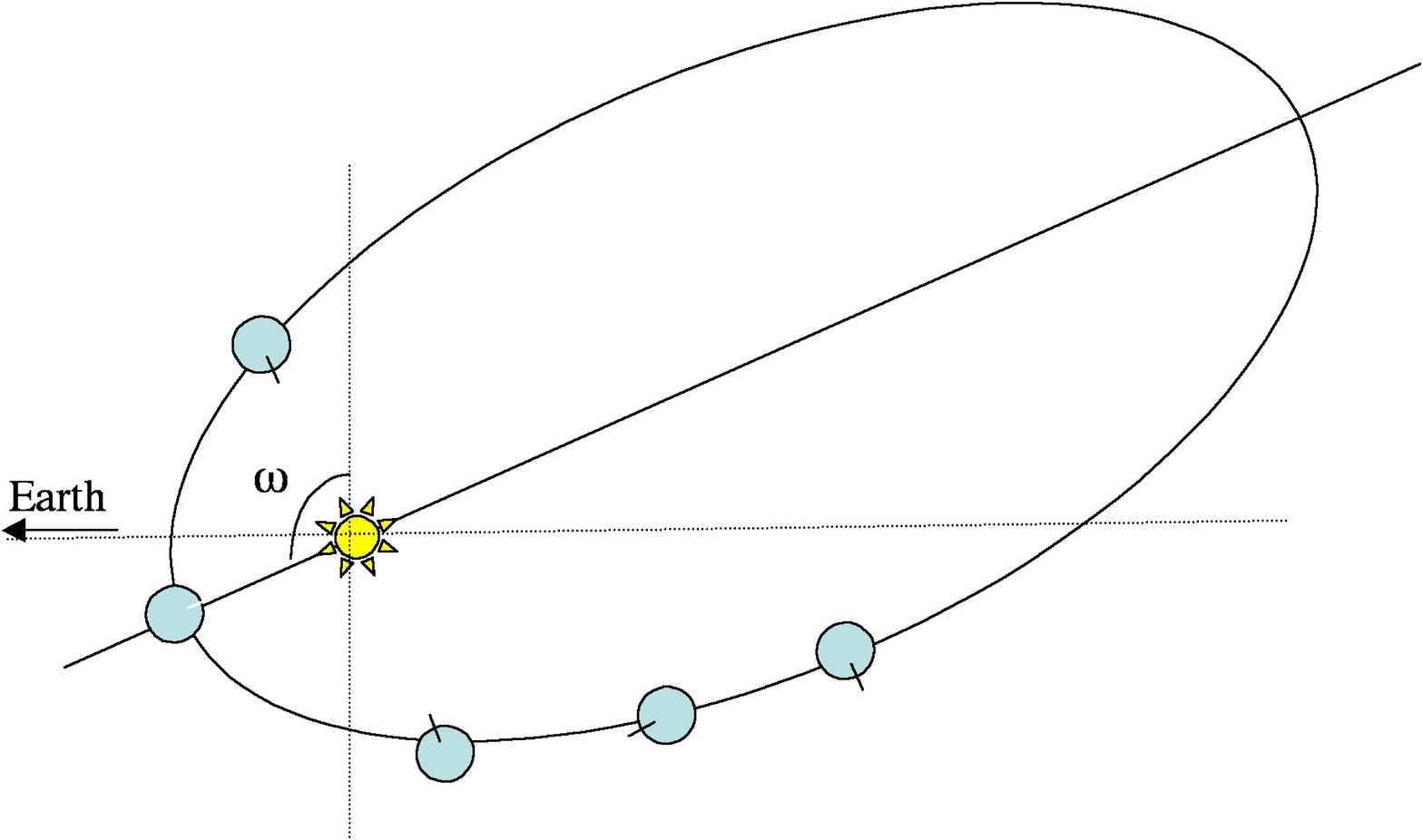}
\caption{The orbit of HD17156b.  Note that the orbit is not drawn to scale and is meant 
for illustration purposes only.  The planet is shown at different positions near periastron 
to illustrate `pseudo-synchronous' rotation.  During the periastron the same side of the 
planet always faces the star, but as the planet moves farther from true anomaly of 90$^o$, 
it begins to rotate asynchronously.  The angle $\omega$ marks the longitude of the 
periastron. \label{fg:orbit}}
\end{figure}

Figure~\ref{fg:orbit} is a crude illustration of the orbit of HD17156b, which also 
shows the orientation of the orbit with respect to an observer on Earth.  The 
transit is offset clockwise from periastron by 31.1 degrees.  The inclination of 
the orbit is $i =$~85.4 degrees \citep{gillon07}, although this figure has been 
revised to 86.5 degrees by \citet{irwin08}.  It is not clear if a secondary eclipse 
exists.  \citet{gillon07} suggest that it should be a partial grazing eclipse, 
but \citet{irwin08} ascribe only a 9.2 \% chance for this to be case.  The probability 
of a full secondary eclipse is even lower at 6.9 \%.  In case secondary eclipse 
does occur, the angular separation between the antitransit and periastron is 
148.9 degrees.    

Simulating the planet's progress on its eccentric orbit is done by varying the irradiation 
conditions continuously as the planet moves along the orbit.  For this purpose, we need to 
work out the true anomaly, which is the angular distance of the planet from periastron, 
and the distance of the planet from the star as a function of time.  Also, as we assume 
that the planet is locked in `pseudo-synchronous' rotation around its axis, we need to account for 
the planet's spin by varying the position of the star in the corotating frame of the planet. 

The `pseudo-synchronous' spin angular velocity is given by \citep{hut81}:

\begin{equation}
\Omega_{sp} = \frac{1 + (15/2) e^2 + (45/8) e^4 + (5/16) e^6}{\lbrack 1 + 3 e^2 + (3/8) e^4 
\rbrack (1-e^2)^{3/2}} \overline{\Omega}
\label{eq:pseudospin}
\end{equation}      
where $e$ is the eccentricity, and $\overline{\Omega}$ is the mean orbital angular 
velocity.  Using $e\sim$0.67 yields $\Omega_{sp}\sim$5.6 $\overline{\Omega}$.  
This means that during one full orbit the planet spins around its axis 4.6 times in the 
star's frame of reference.

In terms of the orbital angular velocity during periastron, $\Omega_p$, the 
`pseudo-synchronous' spin angular velocity can be expressed as:

\begin{equation}
\Omega_{sp} = \frac{1 + (15/2) e^2 + (45/8) e^4 + (5/16) e^6}{\lbrack 1 + 3 e^2 + (3/8) e^4 
\rbrack (1+e)^2} \Omega_p
\label{eq:periastronspin}
\end{equation}  
Again using $e \sim$0.67 yields $\Omega_{sp} \sim$0.818 $\Omega_p$.  In other words, during 
periastron the planet's spin is slower than the orbital angular velocity, and while passing 
the periastron the planet should therefore revolve `backwards' with respect to the star 
compared to normal, faster spin at other parts of the orbit.  This behaviour is a curious feature 
of `pseudo-synchronisation' and it causes a peculiar jitter in the position of the star in the 
planet's sky near periastron.  

The orbital mean anomaly is given by:

\begin{equation}
M( t ) = \frac{2 \pi}{P} (t - T)
\label{eq:mean_anomaly}
\end{equation}
where $P$ is the orbital period and $T$ is the time of periastron.  This is simply the angular 
distance traversed by the planet in time $(t - T)$ if the orbit was circular.  Mean anomaly can 
be related to the eccentric anomaly by using Kepler's second law, which states that the radius 
vector from the host star to the planet sweeps out equal areas in equal time, and the geometric 
properties of an ellipse.  The relation between the mean and eccentric anomalies is given by 
Kepler's equation:
  
\begin{equation}
M = E - e sin (E)
\label{eq:kepler_equation}
\end{equation} 
where $E$ is the eccentric anomaly.  By using the general equation of an ellipse and geometric 
relations, it is straightforward to show that the true anomaly, $\theta$, is related to the 
eccentric anomaly by:

\begin{equation}
cos (\theta) = \frac{cos (E) - e}{1 - e cos (E)}
\label{eq:true_anomaly}
\end{equation}
Once true anomaly is known, it is easy to solve for the orbital distance as a function of time 
by using:

\begin{equation}
d( t ) = \frac{a(1 - e^2)}{1 + e cos \lbrack \theta( t ) \rbrack}
\label{eq:elliptic_distance}
\end{equation}   
where $a$ is the semi-major axis of the orbit.  

Kepler's equation (\ref{eq:kepler_equation}) is a transcendental equation, which cannot be 
solved analytically.  Fortunately, it can be solved easily by using Newton-Raphson iteration.  
For this purpose, the equation is first written as:

\begin{displaymath}
f(E) = E - e sin E - M = 0
\end{displaymath}
If the initial guess at the solution is given by $E_g$, then the progressive corrections to 
this solution are given by:

\begin{displaymath}
E = E_g - \frac{f(E_g)}{f'(E_g)}
\end{displaymath}
where $f'(E_g)$ is the differential of $f$ with respect to E.  The iteration proceeds until 
an acceptable level of convergence is achieved.  In general, the solution to Kepler's equation 
converges very quickly and only a few iterations are required.  Analytically, it can be shown that 
for $e <$0.99 and an initial guess of $\pi$, convergence is guaranteed.  

The change in the local hour angle in time $\delta t$ is given by the numerical difference of 
the spin angular velocity and orbital angular velocity multiplied by $\delta t$:

\begin{equation}
\delta H_a = \lbrack \Omega_{sp} - \Omega( \theta ) \rbrack \delta t
\label{eq:hour_angle}
\end{equation}  
where $\Omega( \theta )$ is the orbital angular velocity, which depends on the true anomaly of 
the planet's position.  Note that near periastron, where $\Omega( \theta )$ is faster than 
$\Omega_{sp}$, $\delta H_a$ can become negative.  

The above suite of equations allows us to model the stellar irradiation on the atmosphere at 
every point along the orbit.  During every time step the model calculates a new value for the 
mean anomaly.  This value is then converted into true anomaly by using equations 
(\ref{eq:kepler_equation}) and (\ref{eq:true_anomaly}).  The distance to the host star is then 
calculated from equation (\ref{eq:elliptic_distance}) and this distance is used to obtain the 
dilution factor for the stellar XUV flux.  Equation (\ref{eq:hour_angle}) is used to calculate 
the position of the star in the planet's sky.  Given that this procedure is repeated every 
timestep, it proceeds surprisingly swiftly and does not add much to overall computation time.  

\begin{figure}
\plotone{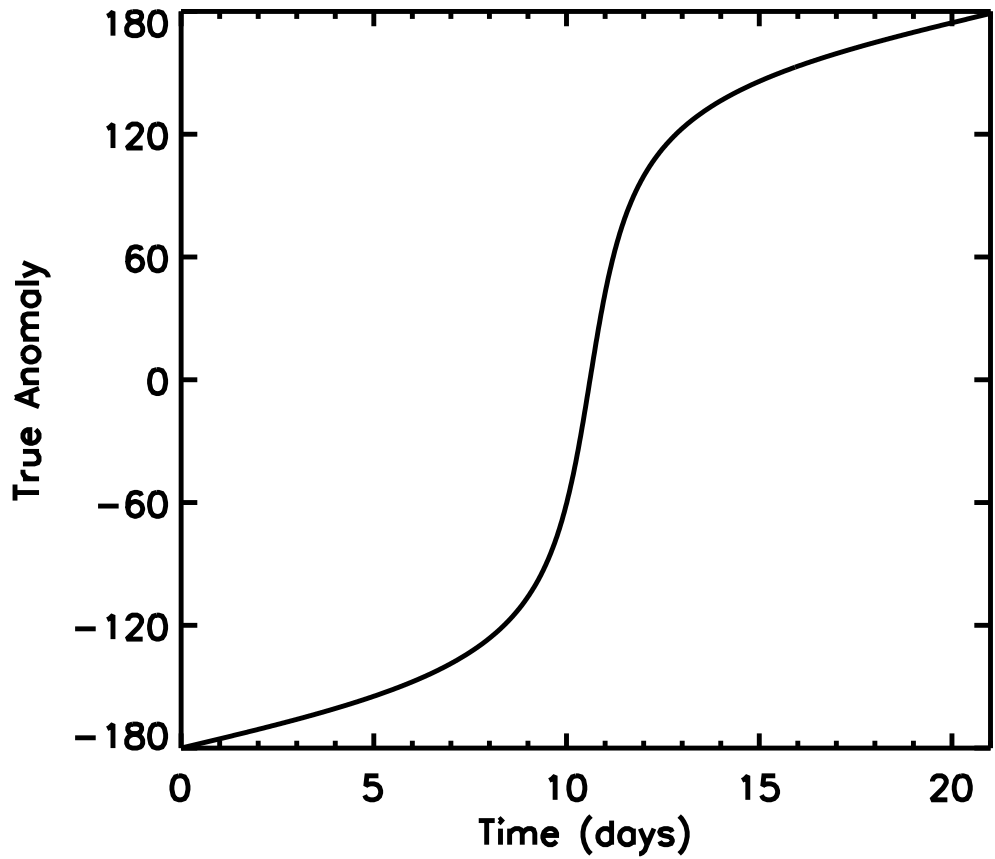}
\caption{Orbital true anomaly versus time (in Earth days).  The orbital period of HD17156b 
is approximately 21.2 Earth days.  Periastron is located at $\theta =$ 0.  The planet 
spends only $\sim$2 days within 0.1 AU, passing through periastron. \label{fg:theta_time}}
\end{figure}

\begin{figure}
\plotone{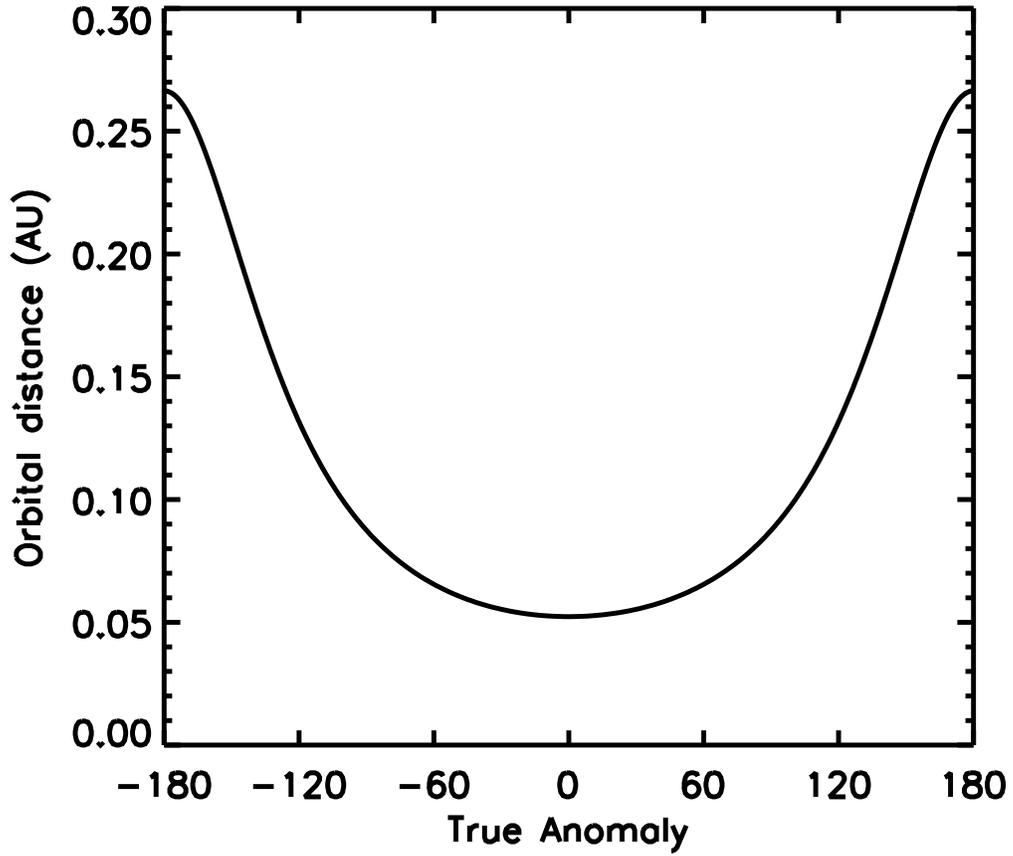}
\caption{Orbital distance (in AU) versus true anomaly.  The orbital period of HD17156b
is approximately 21.2 Earth days.  Periastron is located at $\theta =$ 0. 
\label{fg:dist_theta}}
\end{figure}

\begin{figure}
\plotone{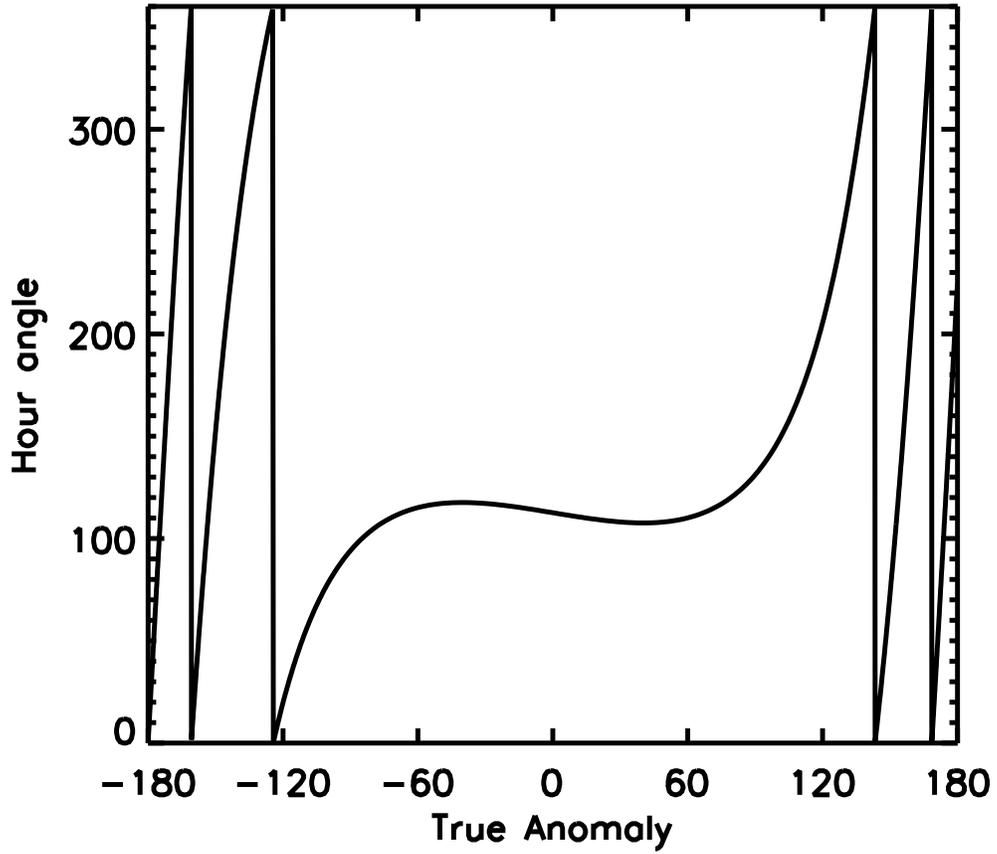}
\caption{Hour angle versus orbital true anomaly.  Here the hour angle traces the angular 
separation of a fixed point on the planet's surface and the substellar point.  Zero and 
360 degrees correspond to the substellar point. \label{fg:ha_theta}}
\end{figure}

Figure~\ref{fg:theta_time} shows the orbital true anomaly versus time for the orbit of 
HD17156b, Figure~\ref{fg:dist_theta} shows the orbital distance versus true anomaly and 
Figure~\ref{fg:ha_theta} shows how the hour angle develops during one orbit.  All simulations 
start from apastron, with $\theta = $~180$^o$, and the local zenith for the hour angle 
calculation is set initially at the substellar point.  The plots illustrate that the orbital 
solution is working as it should and that it makes physical sense.  Also, they demonstrate 
the fact that the orbital angular velocity is faster during periastron, implying that the 
planet spends most of the time completing the `far-side' of the orbit.  

\subsection{Boundary Conditions and Planetary Parameters}

We used planetary and orbital parameters that are listed in Table~\ref{table:parameters} together 
with the other input parameters for the simulations.  We used parameters originally released by 
\citet{gillon07} in their preprint.  These parameters have been subsequently revised, and they 
also differ from those released by \citet{irwin08}.  However, given all the uncertainties 
in the model, and the uncertainties in the published parameters, the differences are not likely 
to be particularly important.  
The numerical solution is evaluated on a grid of 36 evenly spaced longitude points, 31 evenly 
spaced latitude points, and 28 pressure levels ranging from 2 $\mu$bar to 0.04 nbar with a uniform 
spacing of 0.4 scale heights.  The calculations do not extend to the poles and instead, the field 
variable values are interpolated over to the polar latitude circles (31,30,1,2).  We assume that 
the thickness of the lower atmosphere is entirely negligible compared to the radius of the planet, 
and thus place the zero-altitude reference point at our lower boundary at 2 $\mu$bar.

We assume that the lower boundary temperature is constant at T$_0 =$520 K.  This is close to 
the equilibrium temperature of a gas giant with a Bond albedo equal to 0.3 under solar irradiation 
near the apastron of HD17156b's orbit.  Using a constant lower boundary temperature for a planet 
like HD17156b may not be realistic as this temperature is likely to vary along the orbit.  
Unfortunately it is not clear how the temperature varies at the 2 $\mu$bar level as the 
irradiation level changes.  In order to avoid speculation we have chosen to use a constant 
lower boundary temperature for our reference models.  However, as the effect on our model could 
be significant, we also produced simulations where we varied the lower boundary temperature.  
The results are discussed in Section~\ref{sc:varytemp}.  

The initial composition comes from a model of Jupiter's 
auroral region \citep{grodent01} that effectively fixes the mixing ratios of the neutral 
species at the lower boundary of the model.  The lower boundary composition for a planet like 
HD17156b is highly uncertain.  It is not clear if the dominant carbon-bearing molecule in the 
lower atmosphere should be CO or CH$_4$, and there are no constraints on the mixing ratios 
of heavier molecules in general.  Thus we have chosen not to speculate on hydrocarbon or oxygen 
photochemistry, and instead assumed that the mixing ratios of heavier molecules in the 
thermosphere are negligible.  This assumption is not as crude as it may initially appear, 
because due to molecular diffusion the mixing ratios of molecules heavier than He should be 
very small inside our pressure range.       

The mixing ratios of H and He near our lower boundary are also uncertain, which is why our 
first instinct was to use jovian values ($\simÂ$2 $\times$10$^{-4}$ for H, and 0.056 for He).  
However, these values may well be inappropriate for a planet that is so different from 
Jupiter as HD17156b.  For instance, atomic hydrogen in the upper atmosphere is created as 
a by-product of the complex oxygen and hydrocarbon photochemistry, which is powered by 
stellar UV radiation, and is expected to be taking place in the stratosphere-mesosphere 
region of close-in gas giants.    
Calculations of \citet{liang03}, based on such photochemistry, demonstrate that the mixing 
ratio of atomic hydrogen near the $\mu$bar level could be around 1 \% for a close-in giant like 
HD209458b (with a~$=$~0.045 AU).  They indicate that for a variety of chemistries, the 
production of atomic hydrogen is mainly
limited by the strength of the UV flux, which varies 27-fold for HD17156b.  Based on these 
arguments, we expect the mixing ratio of atomic hydrogen to be less than 1 \% for HD17156b 
at the $\mu$bar level because most of the time the planet is farther from the host star 
than HD209458b, and thus it should receive less UV radiation.   

While our model is not sensitive to small variations in 
the mixing ratio of He at the lower boundary, the amount of atomic hydrogen could be crucial because 
it regulates the availability of H$_2$ and thus limits the formation of H$_{3}^{+}$.  
In order to examine the effect of changing the lower boundary mixing ratio of atomic hydrogen, 
we have generated both Jupiter-type models and models where the mixing ratio of atomic hydrogen 
was raised to 1 \% at the lower boundary at the expense of the mixing ratio of H$_2$.         

For all of our simulations, we assume zero winds at the lower boundary.  This point was 
extensively discussed in \citet{koskinen07a}, where we also demonstrated that the inclusion 
of winds and temperature variations at the lower boundary does not affect the conditions 
in the upper atmosphere significantly.  This insensitivity is due to the very strong forcing 
by the stellar XUV radiation in the middle-to upper thermosphere.  As an upper boundary condition, 
we assume that vertical gradients reduce to zero in the two uppermost layers, vertical winds vanish 
at the upper boundary, and that the neutral species are in diffusive equilibrium in the two 
uppermost levels.

\section{Results}
\label{sc:results}

\subsection{Evaporation of the Atmosphere}
\label{subsc:stability}

The stability of the atmosphere against thermally driven hydrodynamic escape is determined 
at the exobase, which is defined as the altitude where an escaping particle undergoes 
approximately one collision within one atmospheric scale height.  A handy measure of stability 
is the thermal escape parameter, given by \citet{hunten73}:

\begin{equation}
\lambda = \frac{G M_p m}{k T_c z_c} = (\frac{v_{esc}}{u})^2
\label{eq:lambda}
\end{equation}
where $m$ is the mass of the escaping constituent, $k$ is Boltzmann's constant, $T_c$ is 
the temperature at the critical level (exobase), and $z_c$ is the altitude of the critical 
level.  This is basically the square of the ratio of the escape velocity to the most probable 
thermal velocity, $u$, at the exobase.  Once the thermal escape parameter becomes less than 
about 1.5 for H or less than about 3.0 for H$_2$, the thermal kinetic energy at the critical 
level becomes greater than the gravitational potential energy and the atmosphere begins 
to escape hydrodynamically.  If $\lambda >$3, Jeans escape dominates. 
Of course, the transition between Jeans escape and hydrodynamic escape is not 
abrupt, and some overlap is to be expected for low values of $\lambda$.  However, at 
$\lambda >$30, even Jeans escape is fairly insignificant. 

\begin{figure}
\plotone{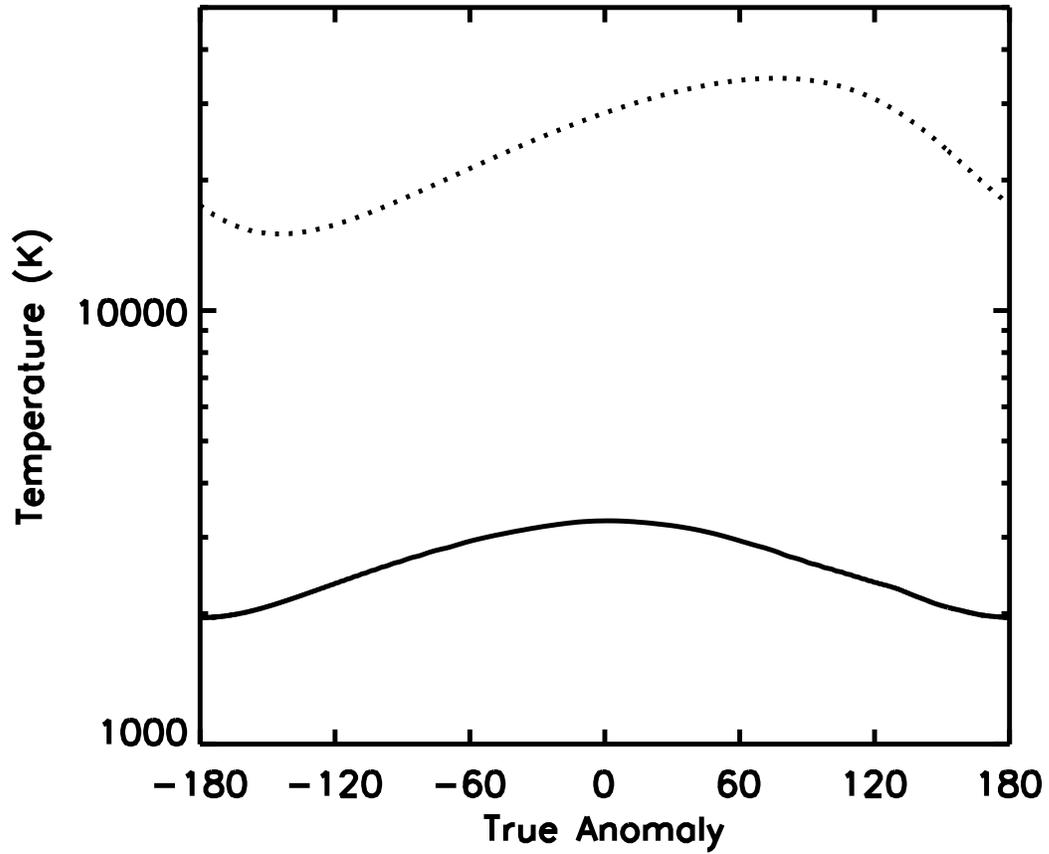}
\caption{Globally averaged upper boundary (at 0.04 nbar) temperatures for two different 
simulations with different lower boundary mixing ratios of atomic hydrogen and radiative 
cooling functions.  The solid line shows temperatures for a model with a cooling function 
of 70-90 \% and the dotted line shows temperatures for a model with a cooling function of 
0.01-0.24 \% (see text). \label{fg:temp_orbit}}
\end{figure}

The escape velocity from HD17156b is 108 kms$^{-1}$, which compares to about 60 kms$^{-1}$ 
for Jupiter.  Thus the temperature in the upper atmosphere of HD17156b must be significantly 
higher for hydrodynamic escape to take place than what is required for a Jupiter-like planet.
Figure~\ref{fg:temp_orbit} shows the globally averaged temperatures for HD17156b at the upper 
boundary of the model versus orbital true anomaly (starting from -180$^o$ at apastron).  
The results are shown for two types of simulations, one with lower boundary mixing ratios 
from \citet{grodent01} (Exo-1) and one with the mixing ratio of atomic hydrogen fixed at 
1 \% at the lower boundary (Exo-2). 

For Exo-1, the exobase temperature is around 2,000 K at apastron, and it rises to over 
3,000 K at periastron.  The temperature peaks around periastron and reaches a minimum near 
apastron.  The temperature difference at apastron between successive orbits is $\sim$0 K, 
indicating that the model repeats the same behaviour during each orbit and is therefore 
in a kind of steady state.  Remarkably, this steady state establishes itself after only 
three simulated orbits.  The upper thermosphere is dominated by molecular hydrogen, and 
the thermal escape parameter varies between 690 during apastron and 410 during periastron, 
rendering the atmosphere stable throughout the whole orbit and even Jeans escape negligible.  
As expected, the thermosphere is relatively thin, with the upper boundary altitude ranging from 
840 km during apastron to 1,100 km during periastron.  

For Exo-2, the exobase temperatures are considerably higher, over 10,000 K throughout the 
orbit.  The minimum temperature is around 15,000 K, reached at $\theta =$~-145$^o$, 
and the maximum temperature is 34,000 K, reached after the periastron at 
$\theta =$~76$^o$.  Atomic hydrogen and H$^+$ are virtually the only species in the 
upper thermosphere, and between the minimum and maximum temperature regions the thermal 
escape parameter ranges from 40 to 15, respectively.  This implies that near periastron, 
Jeans escape could be significant and even some bulk flows are possible.  However, according 
to the criteria based on the thermal escape parameter, the atmosphere should not undergo 
fast hydrodynamic escape at any point along the orbit. 
The upper boundary 
altitude ranges from 13,100 km (1.2 R$_p$) near the temperature minimum to 27,300 km 
(1.4 R$_p$) near the temperature maximum.  The mass loss rate from Exo-2, based on thermal 
Jeans escape, peaks around 10$^6$ gs$^{-1}$ at around $\theta =$76$^o$.  However, this 
mass loss rate is not sustained throughout the orbit, and instead it drops to less 
than 10$^{-2}$ gs$^{-1}$ around apastron. 

The temperature variation between successive apastron passes for Exo-2 is $\sim$130 K after 
fifteen simulated orbits.  This means that the simulations have not reached exact steady state, 
alhough it is reasonable to assume that they are near steady state.  130 K is not a particularly 
significant figure compared to the temperature of the upper boundary (over 15,000 K), 
and the difference gets smaller during each orbit.  Computational constraints mean that 
we cannot run the Exo-2 simulation to exact steady state within a reasonable time period, 
and thus it is possible that the model keeps heating up slowly during each orbit until 
it reaches conditions that would lead to hydrodynamic escape.  Given the rather swift 
passage through the periastron, however, we think that this is unlikely.         

\begin{figure}
\plotone{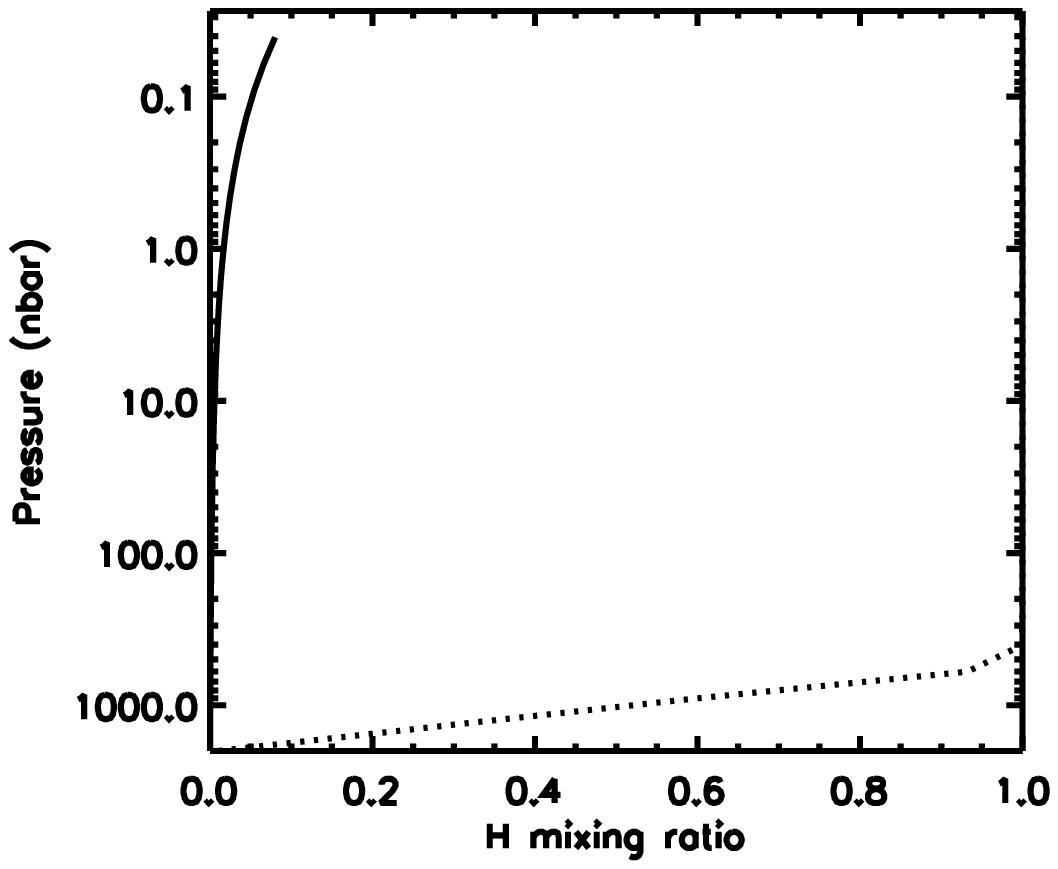}
\caption{The mixing ratio of atomic hydrogen at the substellar point of Exo-1 (solid line) 
and Exo-2 (dotted line) during apastron. \label{fg:hpp}}
\end{figure}

The drastic difference in the thermal characteristics of the two simulations of HD17156b, 
Exo-1 and Exo-2, arises from the differences in the composition of the thermosphere.  
Figure~\ref{fg:hpp} shows the substellar mixing ratios of atomic hydrogen during apastron 
for the two simulations.  For Exo-1, the mixing ratio of atomic hydrogen is relatively low 
at all levels, rising to about 6 \% near the exobase.  At the upper boundary, the maximum 
mixing ratio is about 10 \% near the dawn terminator whereas the minimum is about 5.5 \% 
near the dusk terminator.  For Exo-2, atomic hydrogen becomes the dominant  
species between 1 and 0.1 $\mu$bar (altitude of 50-300 km).  Above that level, atomic 
hydrogen and H$^+$ are virtually the only species in the thermosphere.  
This is because, initially, the higher mixing ratio of atomic hydrogen at the lower boundary 
in the Exo-2 simulation leads to a reduction in the atomic hydrogen concentration 
of the thermosphere through molecular diffusion.  This limits the formation of H$_{3}^{+}$, 
which reduces the significance of radiative cooling compared to the Exo-1 simulation.  
As a result, the temperatures become high enough to dissociate molecular hydrogen thermally, 
and this process quickly converts most of the upper atmosphere into atomic hydrogen that 
is subsequently ionised by the intense XUV radiation.  It should be noted, though, that 
for HD17156b the transition region between the H$_2$ and H atmospheres is located deeper 
in the thermosphere than it would be for a Jupiter-type planet or HD209458b with similar 
lower boundary conditions because of the higher gravity on HD17156b.       

\begin{figure}
\epsscale{.40}
\plotone{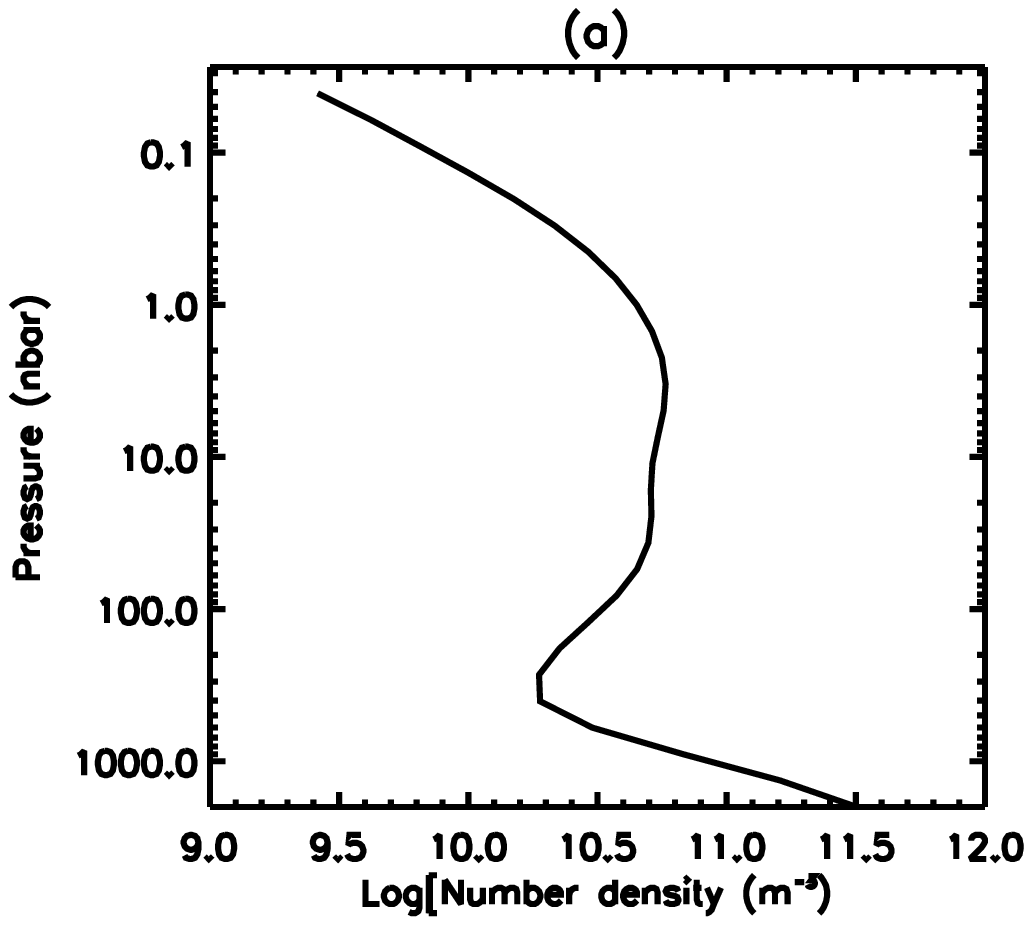}
\plotone{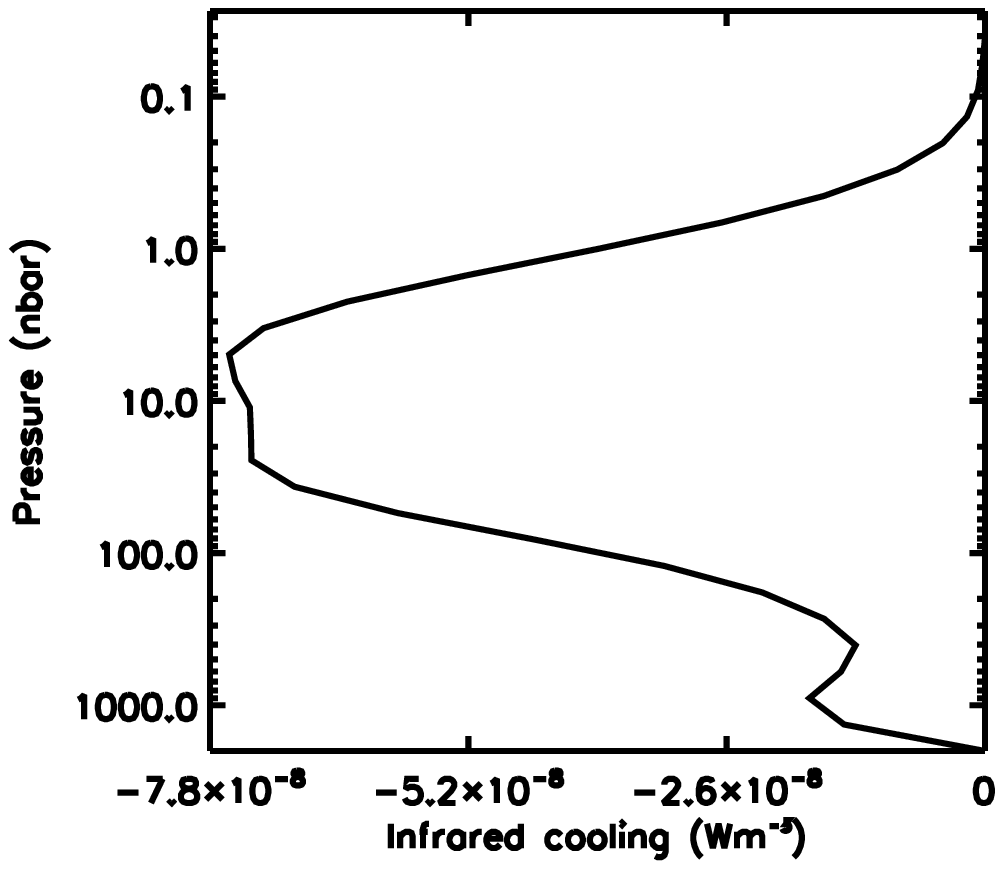}
\plotone{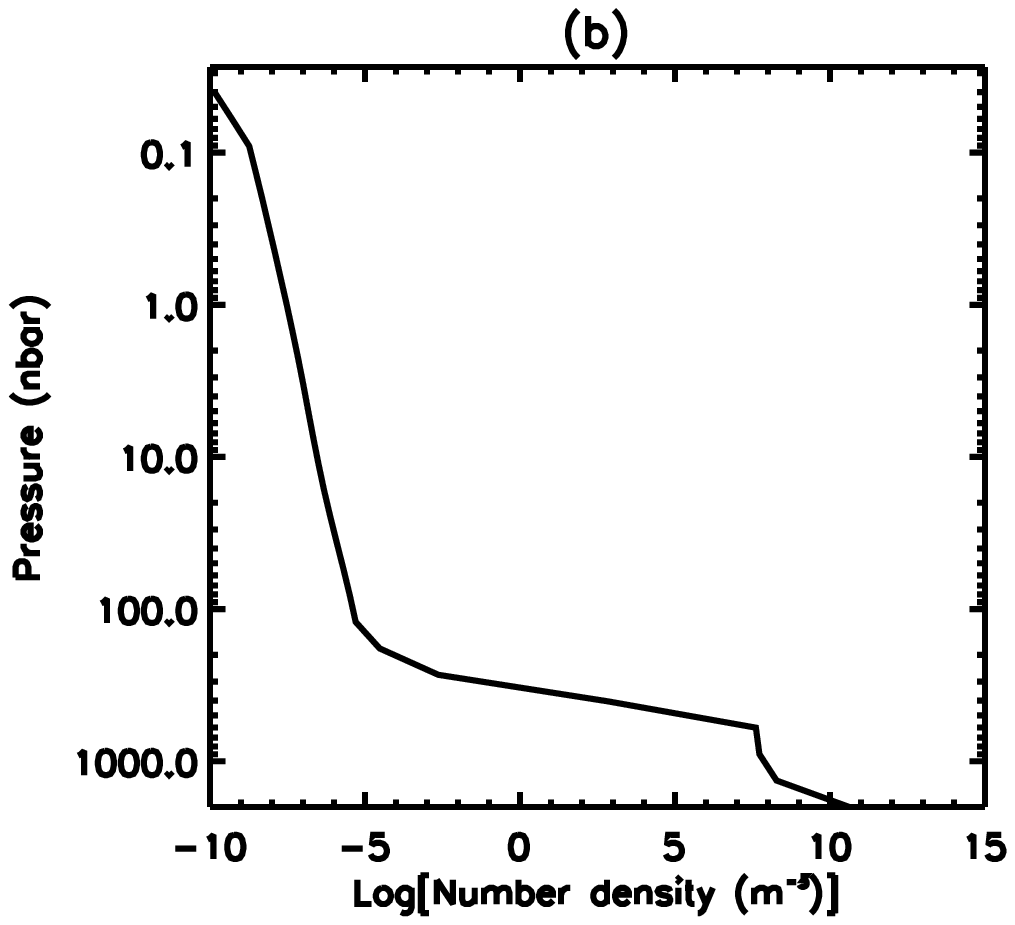}
\plotone{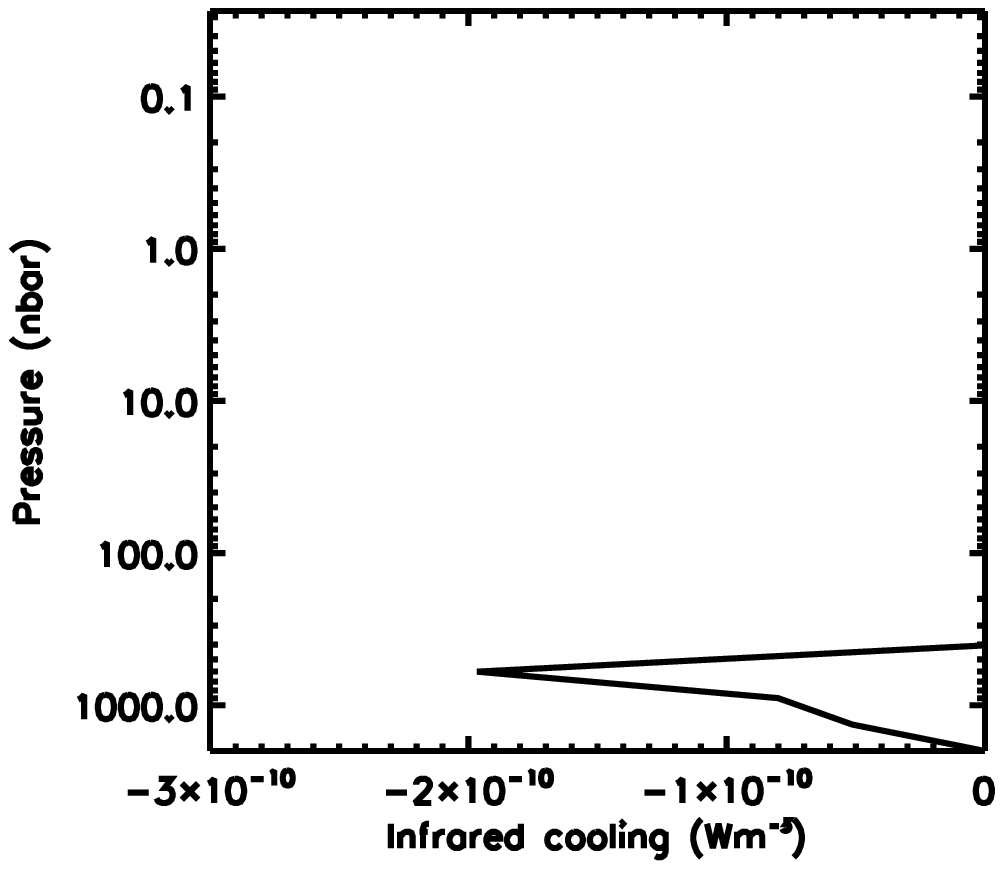}
\caption{H$_{3}^{+}$ densities and the corresponding volume infrared cooling rates 
for (a) Exo-1 and (b) Exo-2 (see text) during apastron at the substellar point. 
\label{fg:h3p_cooling}}
\end{figure}

Figure~\ref{fg:h3p_cooling} contrasts the apastron H$_{3}^{+}$ densities and infrared 
cooling rates at the substellar points of Exo-1 and Exo-2.  The substellar column densities 
of H$_{3}^{+}$ in the two models are 3.2 $\times$10$^{16}$ m$^{-2}$ and 5.0 $\times$10$^{14}$ 
m$^{-2}$, respectively, and in Exo-2 H$_{3}^{+}$ virtually disappears at pressures lower 
than 0.6 $\mu$bar (above 150 km).  This difference in H$_{3}^{+}$ densities makes itself 
evident in the infrared cooling rates, that are prominent for Exo-1 but negligible for 
Exo-2.  Contrasting the total infrared cooling rates to the total XUV heating rate allows 
us to estimate global cooling functions for the models.  Depending on the orbital 
positions, the cooling function for Exo-1 is 72-91 \% while for Exo-2 it is merely 
0.1-0.24 \%.  The cooling function thus depends on the composition and photochemistry of the 
models, and the degree to which H$_2$ is dissociated thermally.

It is unfortunate that the conditions in the thermosphere depend on the lower boundary 
composition, as the mixing ratios there are difficult to constrain.  On the other hand, this 
feature opens up interesting new avenues for observations.  For instance, an observation 
of H Lyman $\alpha$ absorption during transit could be used to constrain the properties of 
the upper atmosphere of HD17156b in the same way as the detection of the extended hydrogen 
cloud around HD209458b \citep{vidalmadjar03} has allowed for the characterisation of that 
planet's atmosphere.  We predict that for the upper atmosphere of HD17156b to be dominated 
by atomic hydrogen, it should extend to more than 1.4 R$_p$ during transit near periastron.  
If an extended hydrogen atmosphere is not detected, then infrared cooling must be taking place 
and the thermosphere is dominated by H$_2$.  Thus, even if H$_{3}^{+}$ emissions are too 
faint to be detected directly, we could infer their existence indirectly and also constrain 
the mixing ratios of H and H$_2$ in the lower atmosphere. 

\subsection{Circulation}
\label{subsc:circulation}

\begin{figure}
\epsscale{.75}
\plotone{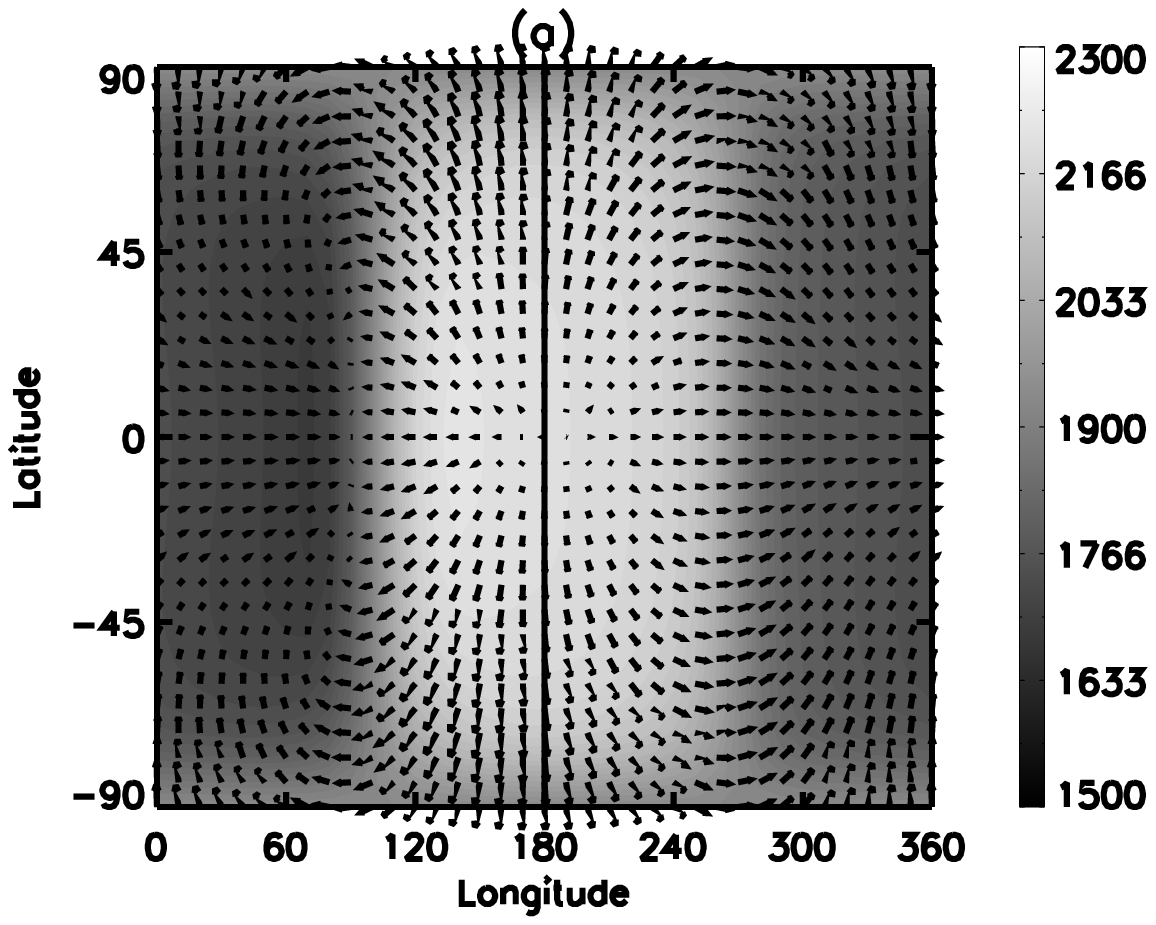}
\plotone{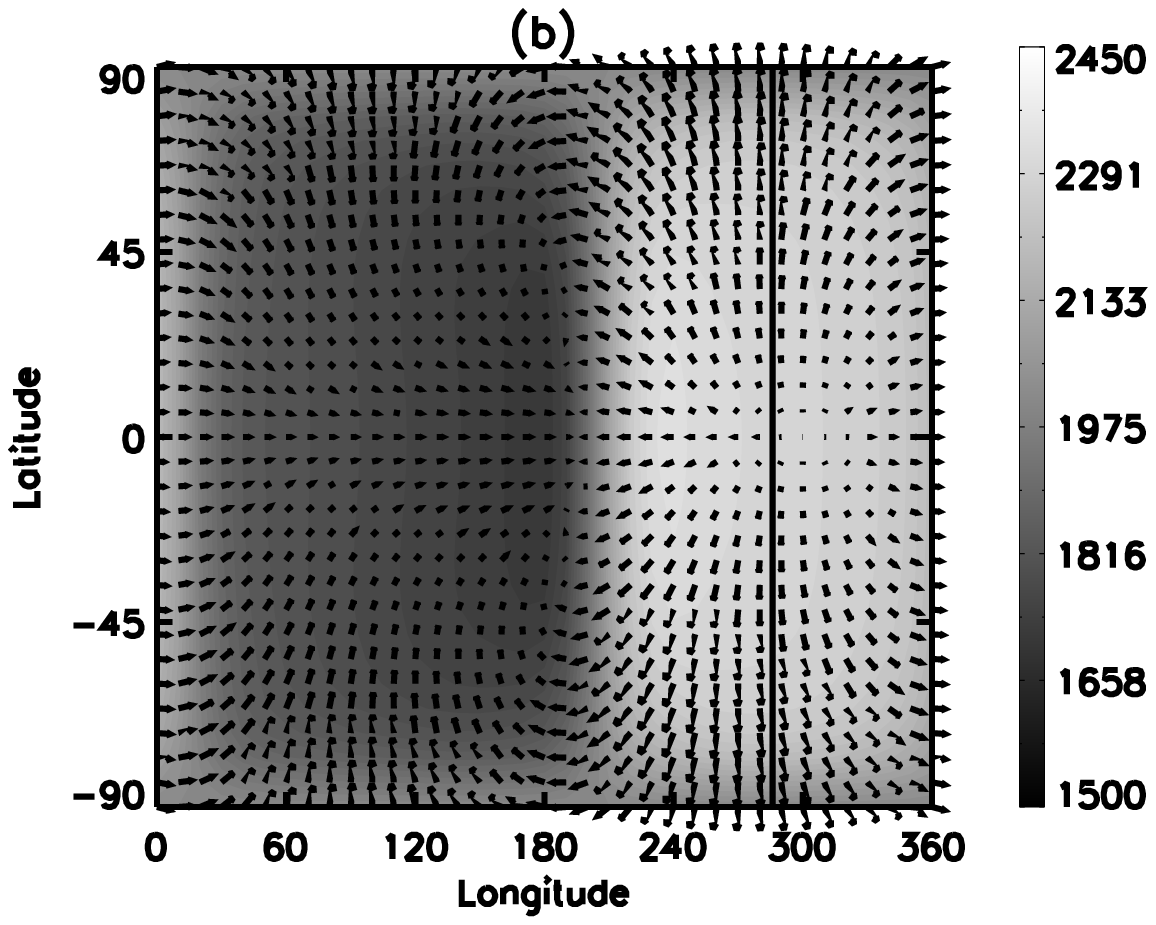}
\plotone{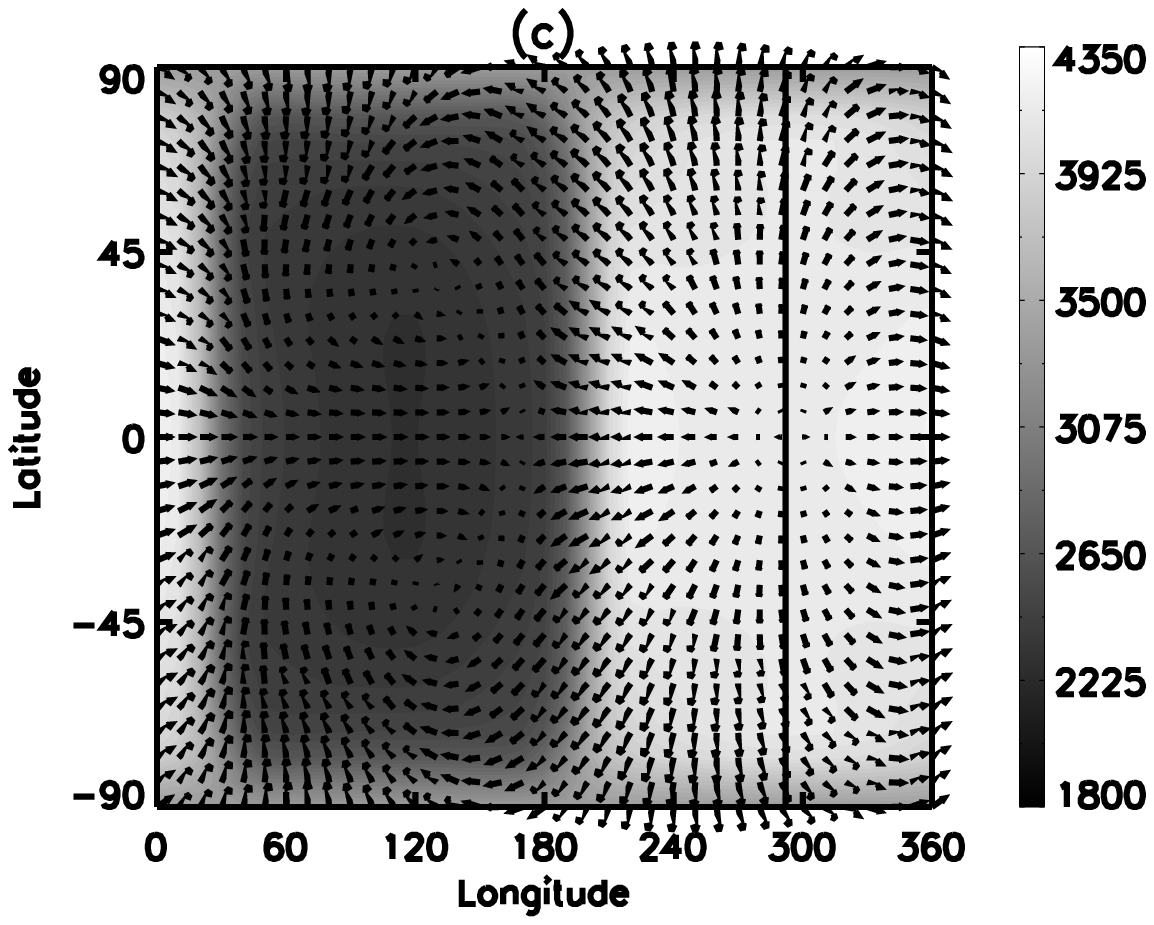}
\plotone{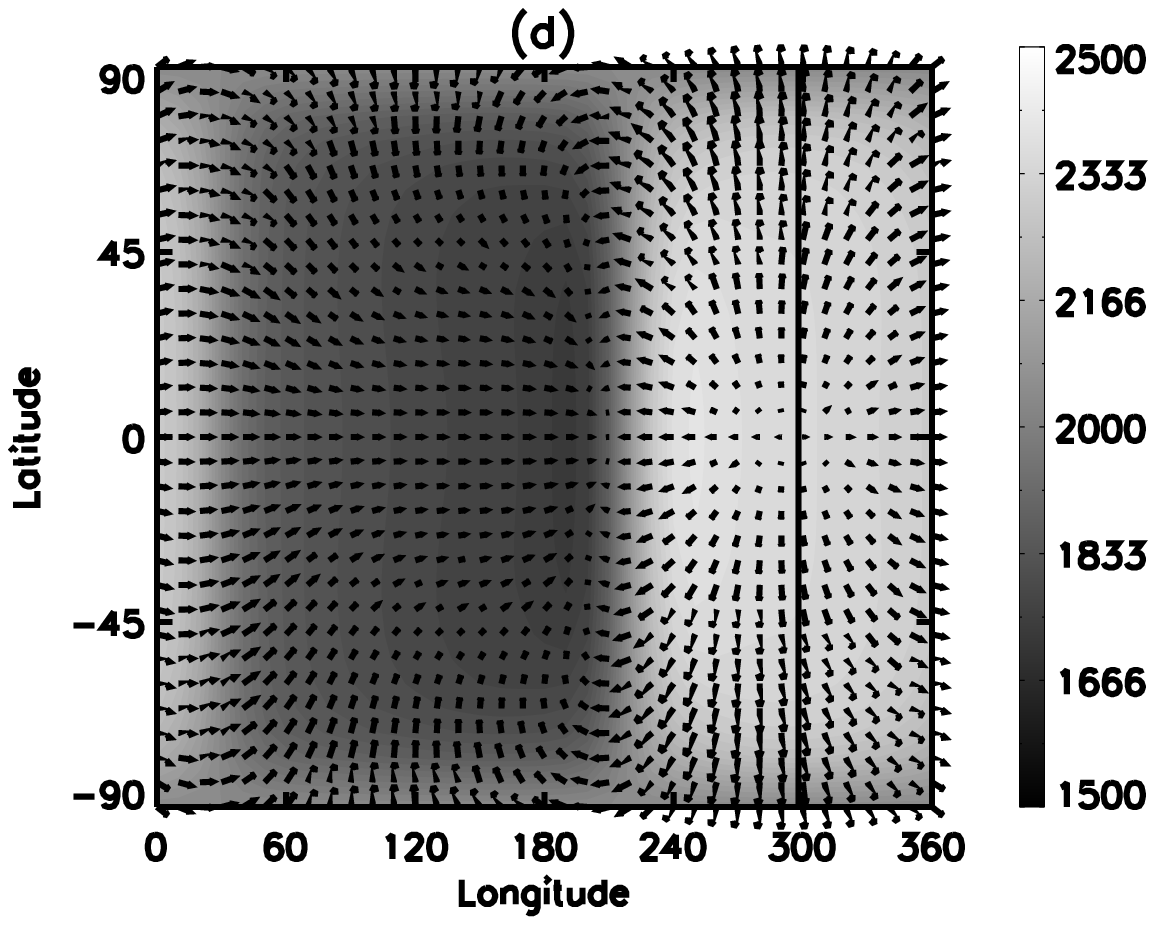}
\caption{Temperatures and winds at the upper boundary of Exo-1 at (a) apastron, 
(b) $\theta =$~-153$^o$, (c) periastron, and (d) $\theta =$~153$^o$.  The 
substellar longitude is marked in each plot with a vertical line.
\label{fg:ex01twtop}}
\end{figure}

Figure~\ref{fg:ex01twtop} shows the temperature and circulation at the upper boundary 
of Exo-1 at four different orbital positions: during apastron, a quarter orbit after apastron 
($\theta =$~-153$^o$), during periastron, and a quarter orbit after periastron 
($\theta =$~153$^o$).  If secondary eclipse takes place, the fourth orbital 
position we have considered here is near (although not exactly at) the longitude of 
the antitransit, wheras the periastron models are a fairly good representation of 
the transit conditions.   

At all four orbital positions, the dayside is clearly warmer than the night side.  
At apastron, the dayside temperature is 2,200-2,300 K, and the temperature peaks 
at the equator near dawn.  The night side temperature, with a minimum close to 
the dawn terminator, is 1,650-1,750 K.  The diurnal temperature difference is 
thus 500-600 K.  
The horizontal winds originate in the dayside, and blow accross the terminator to 
the night side.  In the night side, the Coriolis force and the geopotential 
gradients drive the eastward wind towards the equator, into a stream that flows 
across the antistellar point and faces the westward wind from the dayside near the 
dawn terminator.  Also in the night side, there are two high-latitude vortices 
that direct the easterly wind from the dayside into the westerly equatorial jet.
The maximum zonal wind speed is 1.1-1.3 kms$^{-1}$ at high latitudes.   
Convection is directed upwards in the dayside, accompanied with adiabatic expansion 
of the atmosphere, and downwards in the night side, accompanied with adiabatic 
contraction of the atmosphere.  The maximum vertical flow speed is only a 
few ms$^{-1}$. 
The thermal escape parameter for the average atmosphere at the upper boundary is 689, 
indicating negligible thermal evaporation.  
An escaping particle undergoes approximately 5 collisions within one scale 
height above the upper boundary, implying that the 0.04 nbar level is  
slightly below the exobase.

At $\theta =$~-153 degrees the distance of the planet from the star is 0.22 AU.  
The upper boundary temperature and wind pattern is qualitatively identical to the 
apastron model.  However, with added heating, the dayside temperatures are 
higher, at 2,300-2,450 K.  The night side temperature is 1,650-1,850 K, implying 
a diurnal temperature difference of 600-700 K.  The maximum zonal wind speed 
is also slightly higher at 1.3-1.6 kms$^{-1}$.  
The thermal escape parameter is 659, indicating that thermal evaporation is 
insignificant.  

At periastron, where the orbital distance is 0.052 AU, the near-synchronisation of the 
planet's spin leads to a more pronounced diurnal temperature difference.  Between 
$\theta =$~-90$^o$ and $\theta =$~90$^o$, only one hemisphere of the planet is 
exposed to stellar irradiation.  The dayside temperature during periastron is around 
4,350 K, while the night side temperature is around 2,200 K.  Thus the diurnal temperature 
difference is over 2,000 K.  Large-scale circulation is qualitatively similar to 
other orbital positions, but the winds are faster, with maximum speeds of 
2.6-2.8 kms$^{-1}$.  Vertical advection is also more rigorous, and the maximum vertical wind 
speed is 8-15 ms$^{-1}$.  However, despite the added heating, the thermal escape 
parameter is 411, and evaporation is still slow.  The enhanced XUV heating is 
balanced by effective H$_{3}^{+}$ cooling, and before the thermosphere has time 
to heat up significantly, the planet moves away from periastron. 
At $\theta =$~153$^o$, the upper boundary is almost identical to the 
$\theta =$~-153$^o$ model.  There are some very slight differences between these 
models, but those are mostly limited to the lower thermosphere.  

\begin{figure}
\plotone{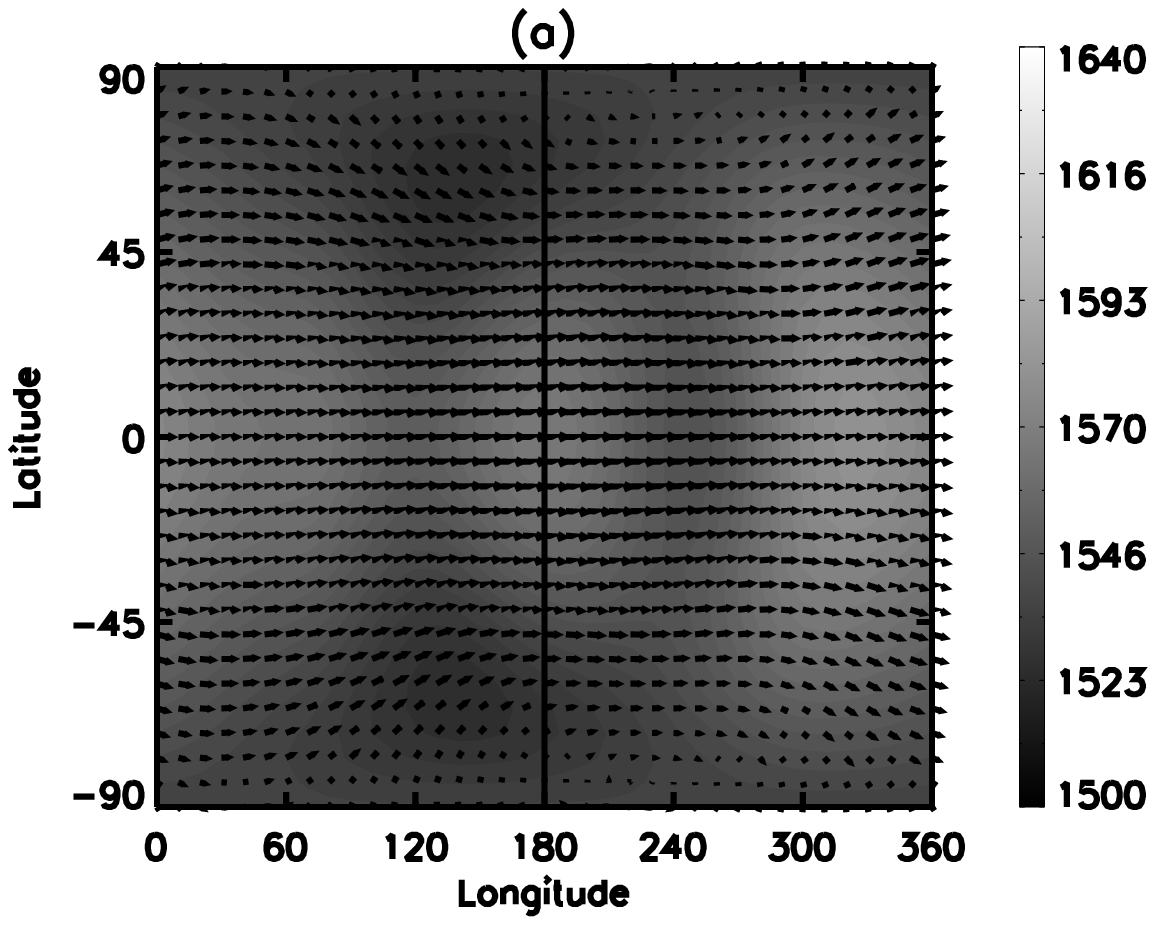}
\plotone{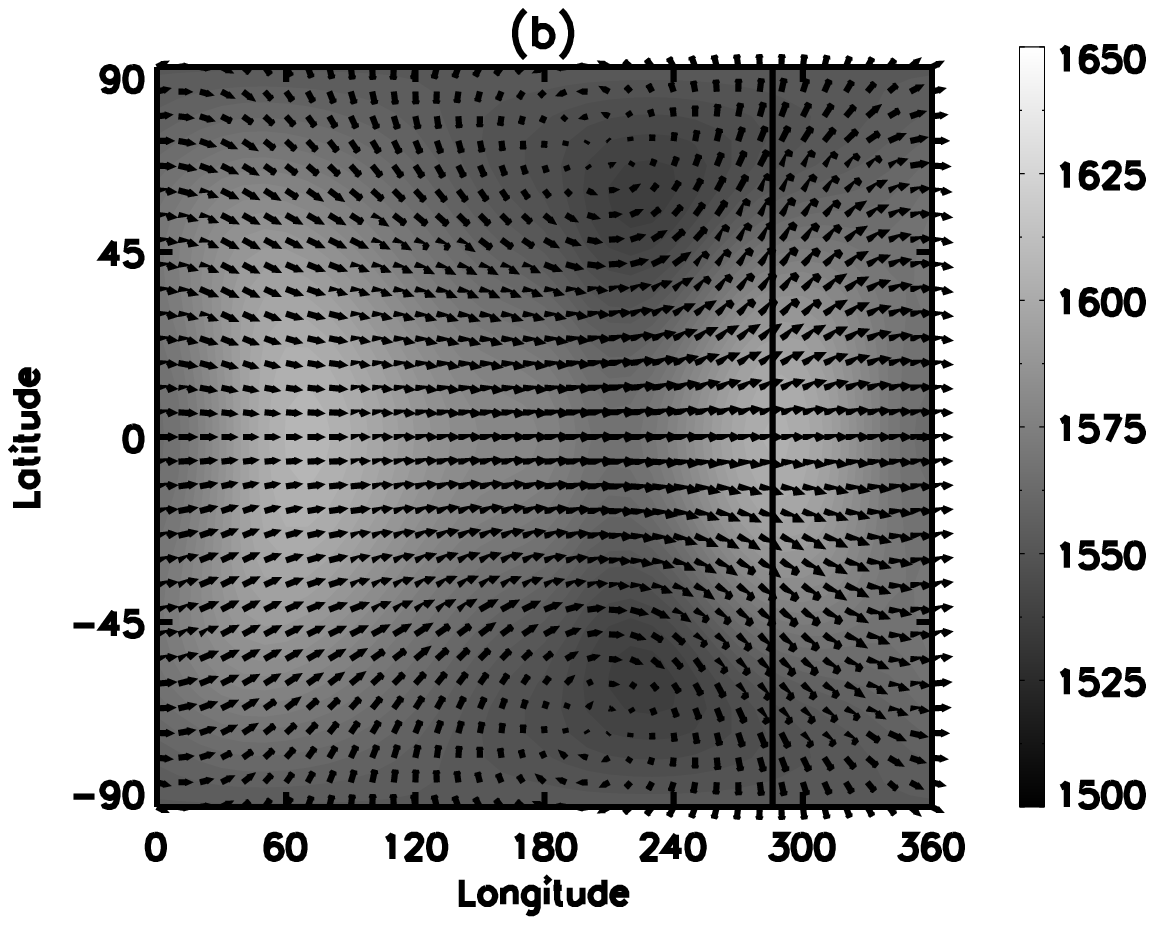}
\plotone{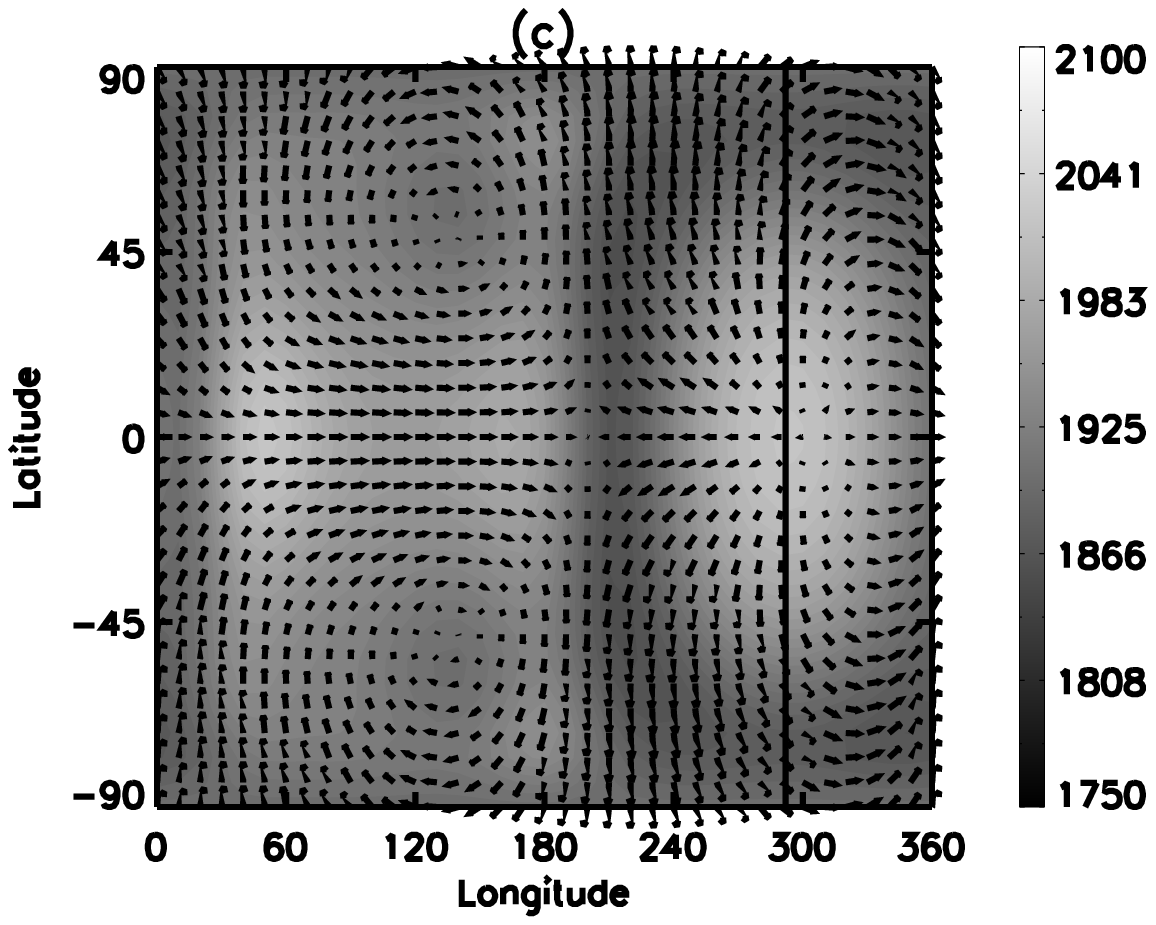}
\plotone{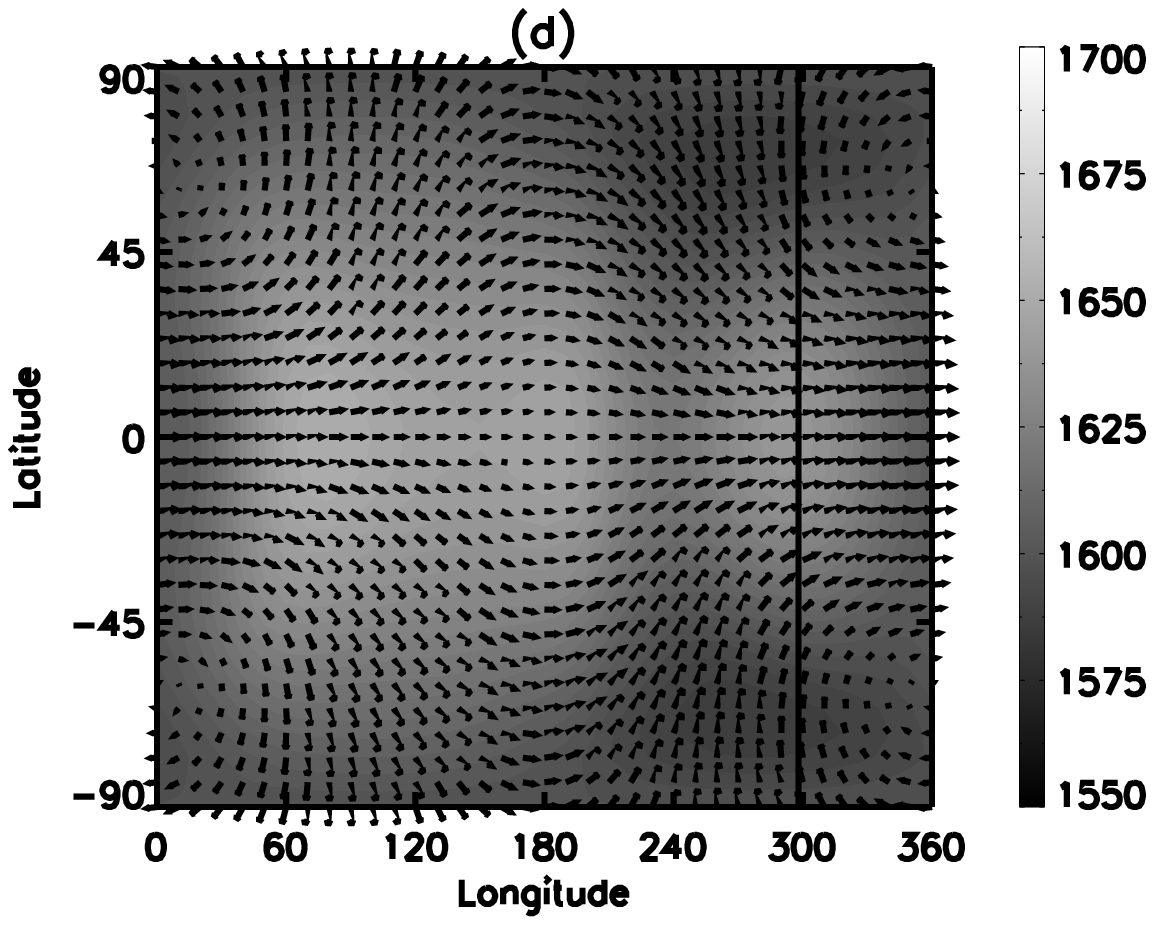}
\caption{Temperatures and winds near the 55 nbar level of Exo-1 at (a) apastron,
(b) $\theta =$~-153$^o$, (c) periastron, and (d) $\theta =$~153$^o$.  The 
substellar longitude is marked in each plot with a vertical line.
\label{fg:ex01twbm}}
\end{figure}

Deeper in the thermosphere, the situation is different.  Figure~\ref{fg:ex01twbm} 
shows the temperature and circulation near 55 nbar (altitude of 250-300 km) for Exo-1 
at the same orbital positions as above.  In all cases, the temperature is nearly 
uniform.  With the exception of periastron, the circulation is characterised by 
a broad eastward, circumplanetary jet. 
During apastron, the temperature varies between 1,530 and 1,580 K, and the equator 
is everywhere slightly warmer than its surroundings, with temperature peaking in the 
night side.  The wind speed in the equatorial jet ranges from 160 to over 220 ms$^{-1}$.  
Cyclonic polar vortices circle around the minimum temperature regions near the poles.
In general, the circulation in the lower thermosphere is characteristic of 
Coriolis-driven dynamics, that smoothes out diurnal temperature variations.  

At $\theta =$~-153$^o$, the temperature varies between 1,540 and 1,610 K.  
The equator is still warmer than its surroundings, and there are two temperature 
peaks, one near the substellar point and one in the night side.  The temperature 
minima are again located near the poles near the dawn terminator, and the cyclonic polar 
vortices are centred on those.  In the equatorial jet, the wind speed ranges from 100 to 
220 ms$^{-1}$.

During periastron the horizontal temperature variations are more pronounced, with 
the temperature ranging from 1,840 to 2,020 K, but the details of the temperature 
distribution are confusing.  The warm substellar region is surrounded by a cooler 
ring along the terminator, but the night side is almost as warm as the dayside.  
The winds diverge from the substellar point, blowing towards the night side.  
An eastward jet develops in the night side, but it encounters the easterly wind at 
dawn, and cannot encircle the planet at this pressure level.  The two high-latitude 
vortices direct easterly flows into the equatorial jet.  The maximum wind speed is 
around 800 ms$^{-1}$.  The circulation is qualitatively similar 
to the topside circulation, and this must be a consequence of the near-synchronisation 
during periastron.   

At $\theta =$~153$^o$, the temperature ranges from 1,580 to 1,660 K. 
Overall, the lower thermosphere is slightly warmer than it was at the symmetric 
position at $\theta =$~-153$^o$, and it is in the process of cooling down after 
the periastron passage.  
The circulation is again characterised by the circumplanetary jet and the polar 
vortices.  The wind speed in the equatorial jet ranges from 100 to 400 ms$^{-1}$.  

\begin{figure}
\plotone{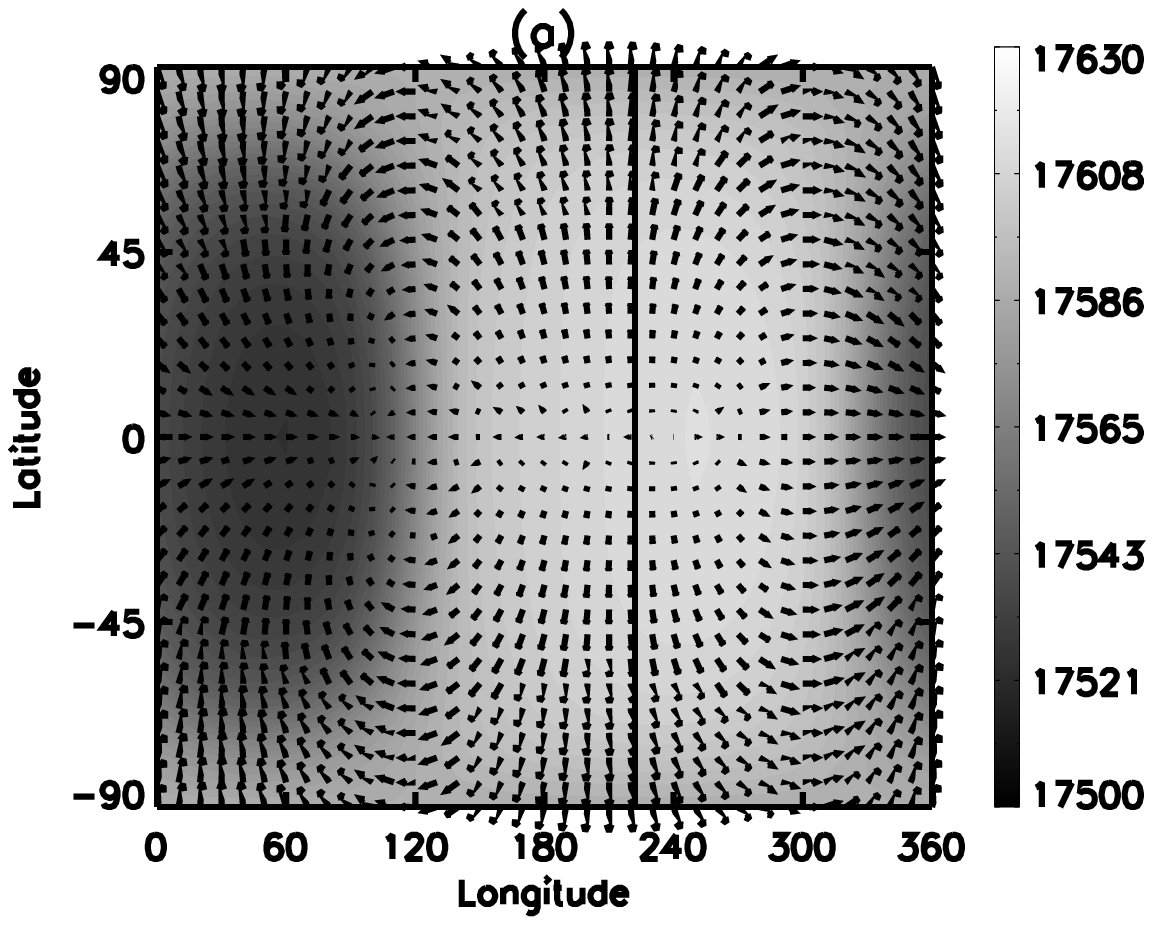}
\plotone{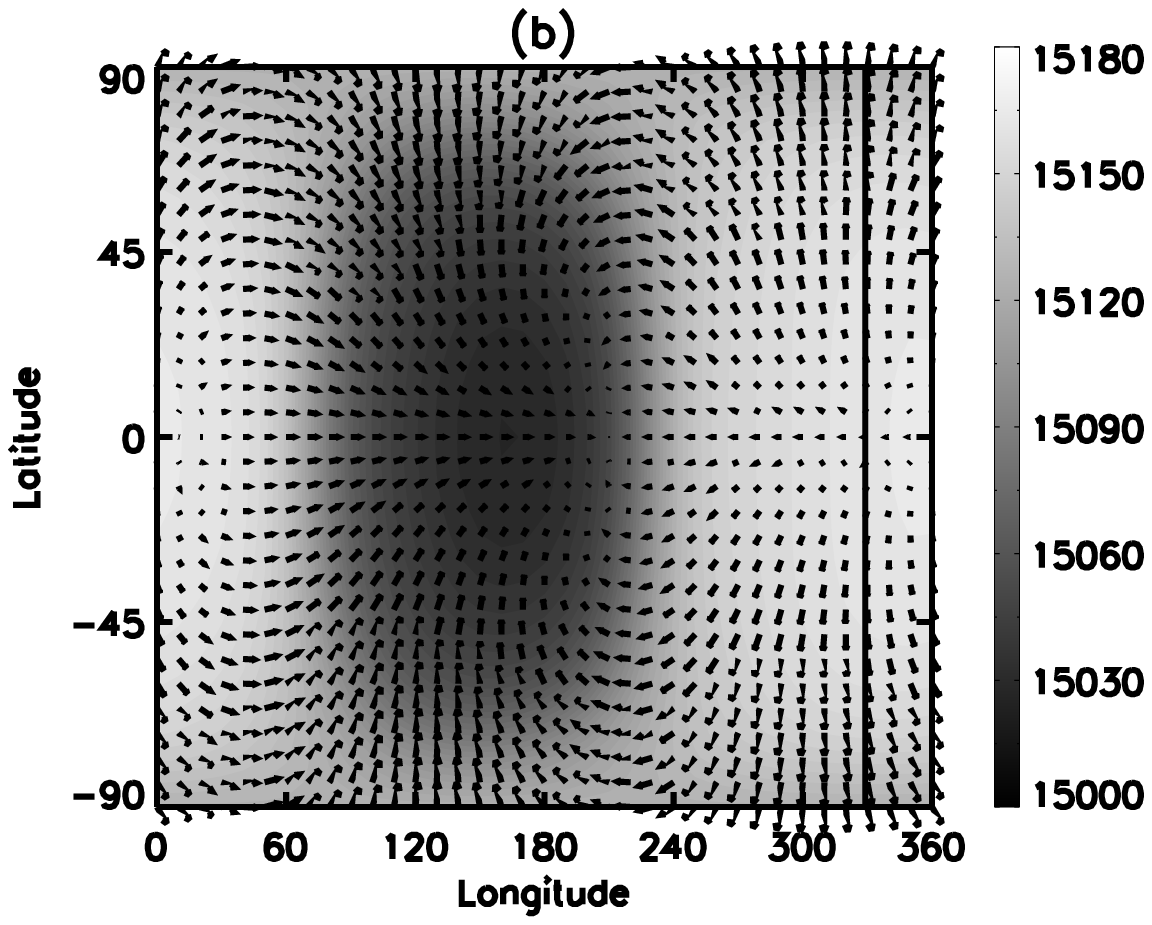}
\plotone{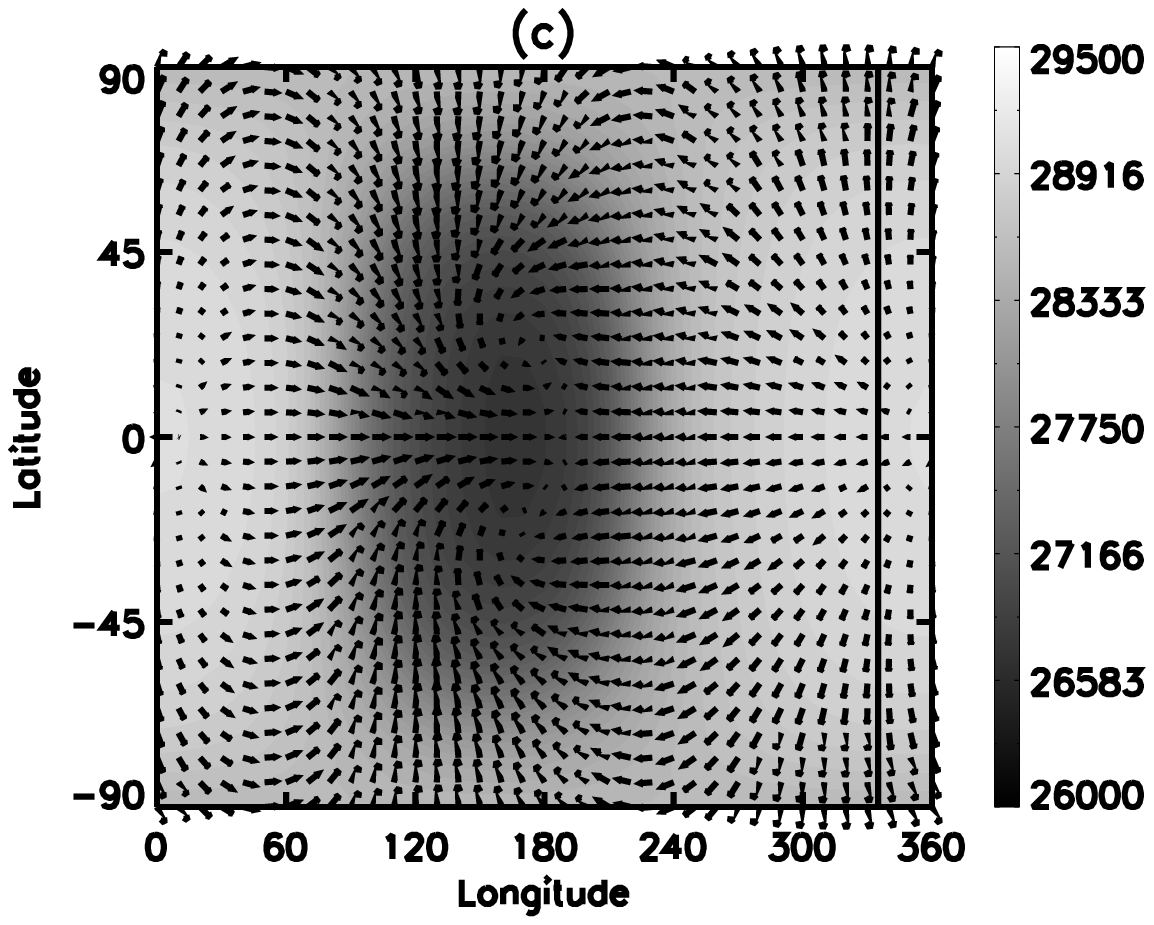}
\plotone{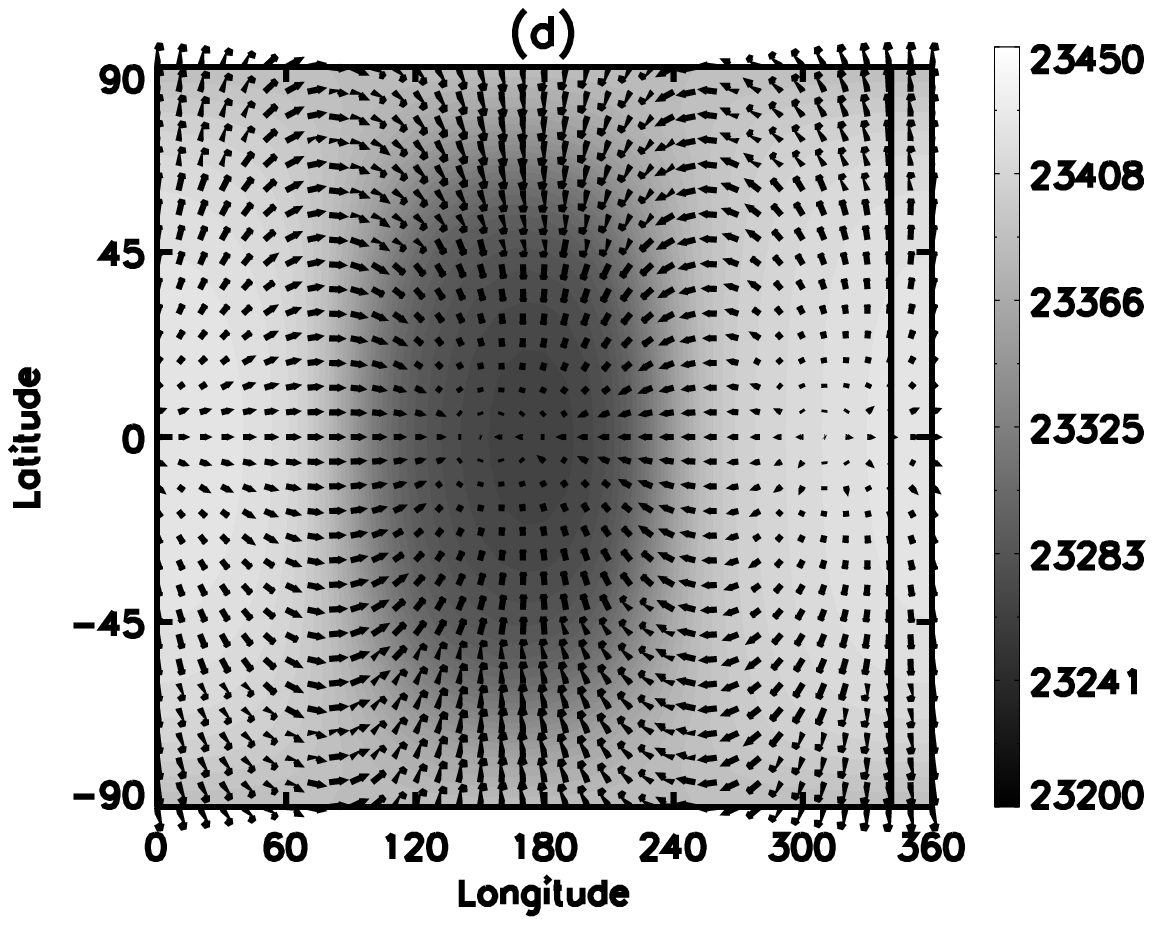}
\caption{Temperatures and winds at the upper boundary of Exo-2 at (a) apastron,
(b) $\theta =$~-153$^o$, (c) periastron, and (d) $\theta =$~153$^o$.  The
substellar longitude is marked in each plot with a vertical line.
\label{fg:ex02twtop}}
\end{figure}

For comparison, Figure~\ref{fg:ex02twtop} shows the temperature and circulation at the 
upper boundary of Exo-2 for the same orbital positions.  
The horizontal uniformity of the temperature in these simulations is remarkable, and 
the wind speeds are very slow.  During apastron, the dayside temperature is 
17,620-17,630 K, while the night side temperature is around 17,530 K.  Thus the diurnal 
temperature difference is only of the order of 100 K.  The winds blow from 
the dayside to the night side, and in the night side some of the eastward flow 
is directed into an equatorial jet that faces the easterly wind from the dayside 
near dawn.  High-latitude vortices turn some of the easterly flow around, and 
direct it into the equatorial jet.  In the dayside, the temperature peak is shifted 
slightly eastward from the substellar point by circulation and rotation.  
The winds are slow, with a maximum speed around 200 ms$^{-1}$.  
An escaping particle suffers $\sim$14 collisions within one scale height at the 
upper boundary, indicating that the 0.04 nbar level is located below the exobase.  
However, the P-T profile is isothermal at pressures lower than $\sim$0.7 nbar, 
and thus the temperature at the critical level should be the same as at our 
upper boundary.  The thermal escape parameter at this boundary is 33, and the 
mass loss rate based on Jeans escape is $\sim$10$^{-2}$ gs$^{-1}$.

At $\theta =$~-153$^o$, the model is generally cooler than at apastron.  The 
dayside temperature is around 15,180 K, while the night side temperature is around 
15,030 K.  The fact that the model cools down despite increasing XUV fluxes 
implies that a slow cooling process takes place after periastron that continues 
even after apastron.  The circulation is qualitatively identical to the apastron 
model, but the maximum wind speed is slightly higher at 340 ms$^{-1}$.  The 
thermal escape parameter increases to $\sim$39 at the upper boundary, and the 
mass loss rate is $\sim$10$^{-4}$ gs$^{-1}$. 

Moving from $\theta =$ -153 to periastron the model is heated by the steadily increasing 
XUV flux.  Between these two positions, the flux is multiplied by a factor of 18.  
During periastron, the dayside temperature is 29,200-29,500 K, with a maximum on the 
equator, shifted slightly eastward from the substellar point.  The night side temperature 
is around 26,800 K, implying that the diurnal temperature difference near transit should 
be of the order of 2,500 K.  This implies that during periastron the XUV flux drives the 
model, instead of dynamics and slow cooling.  The winds blow from the dayside to the 
night side, converging near the antistellar point, with a maximum speed of 
2.8-3.2 kms$^{-1}$.  The thermal escape parameter is 19, which indicates that the atmosphere 
is still in the thermal Jeans escape regime.  The mass loss rate is $\sim$10$^4$ gs$^{-1}$. 

By the time the planet reaches $\theta =$~153 degrees, the thermosphere has started cooling down.  
At this point, the dayside temperature is around 23,430 K and the night side temperature is around 
23,270 K.  Thus the diurnal temperature difference is of the order of 150 K.  As the upper 
thermosphere slowly cools, energy is advected around the planet and this mixing smoothes out 
the diurnal temperature difference that develops during periastron.  Curiously, the day-night 
circulation is still symmetric about an axis connecting the substellar point to the antistellar 
point.  The winds have slowed down to less than 200 ms$^{-1}$, and the thermal escape parameter 
has increased to $\sim$23 at the upper boundary.  The mass loss rate is $\sim$2~$\times$10$^3$ 
gs$^{-1}$.  

\subsection{Thermal Structure and Energy Balance}
\label{subsc:thermal_energy}

\begin{figure}
\epsscale{1.0}
\plotone{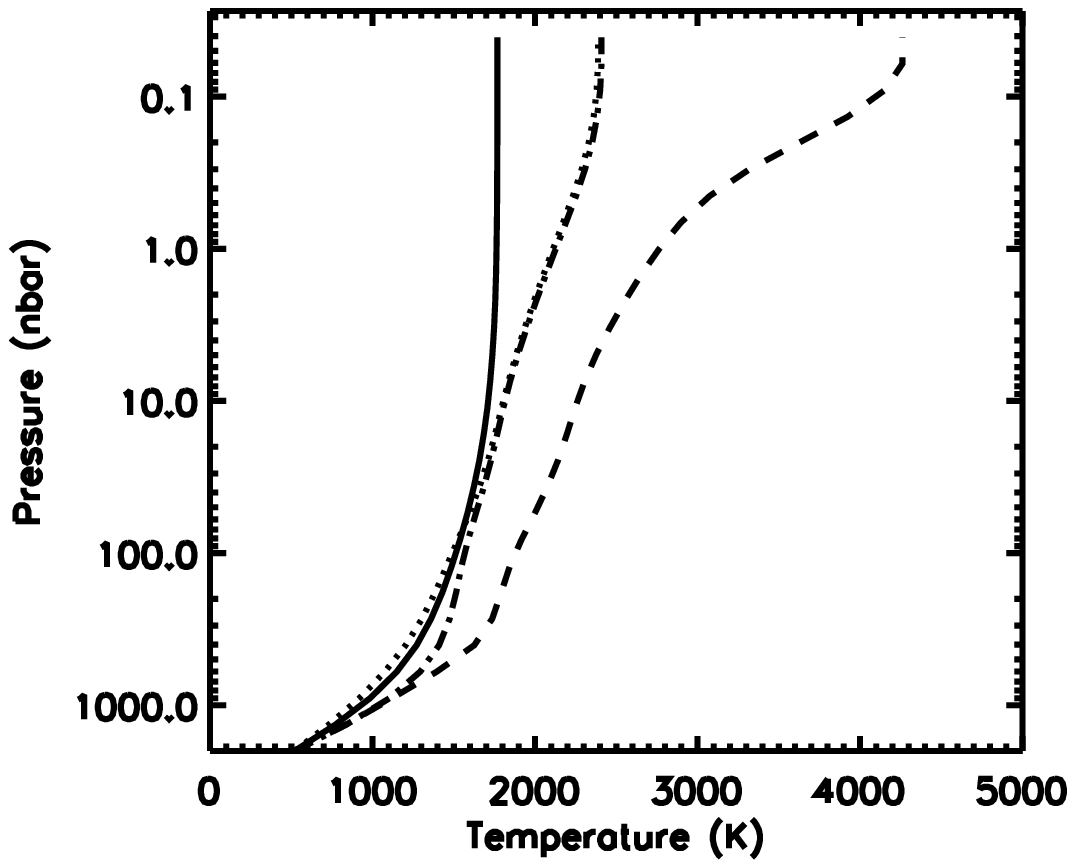}
\caption{Substellar P-T profiles for the Exo-1 simulations at apastron (solid line), 
$\theta =$~-153$^o$ (dotted line), periastron (dashed line), and $\theta =$~153$^o$ 
(dash-dotted line). \label{fg:ex01pt}}
\end{figure}

Figure~\ref{fg:ex01pt} shows the substellar P-T profiles for the Exo-1 series at the four 
orbital positions discussed in the previous section.  Outside periastron the P-T profiles 
are isothermal at pressures lower than about 0.2 nbar.  Generally the temperature increases 
with altitude, and the gradient is steepest in the lower thermosphere at pressures higher 
than 0.3 $\mu$bar.  Overall, the temperatures are highest during periastron, and lowest 
during apastron.  At $\theta =$~-153$^o$ and at $\theta =$~153$^o$ the P-T profiles are 
identical in the upper thermosphere, but the lower thermosphere is warmer after periastron.  
During periastron the temperature rises sharply with altitude towards the upper 
boundary due to added XUV heating.   

\begin{figure}
\epsscale{.45}
\plotone{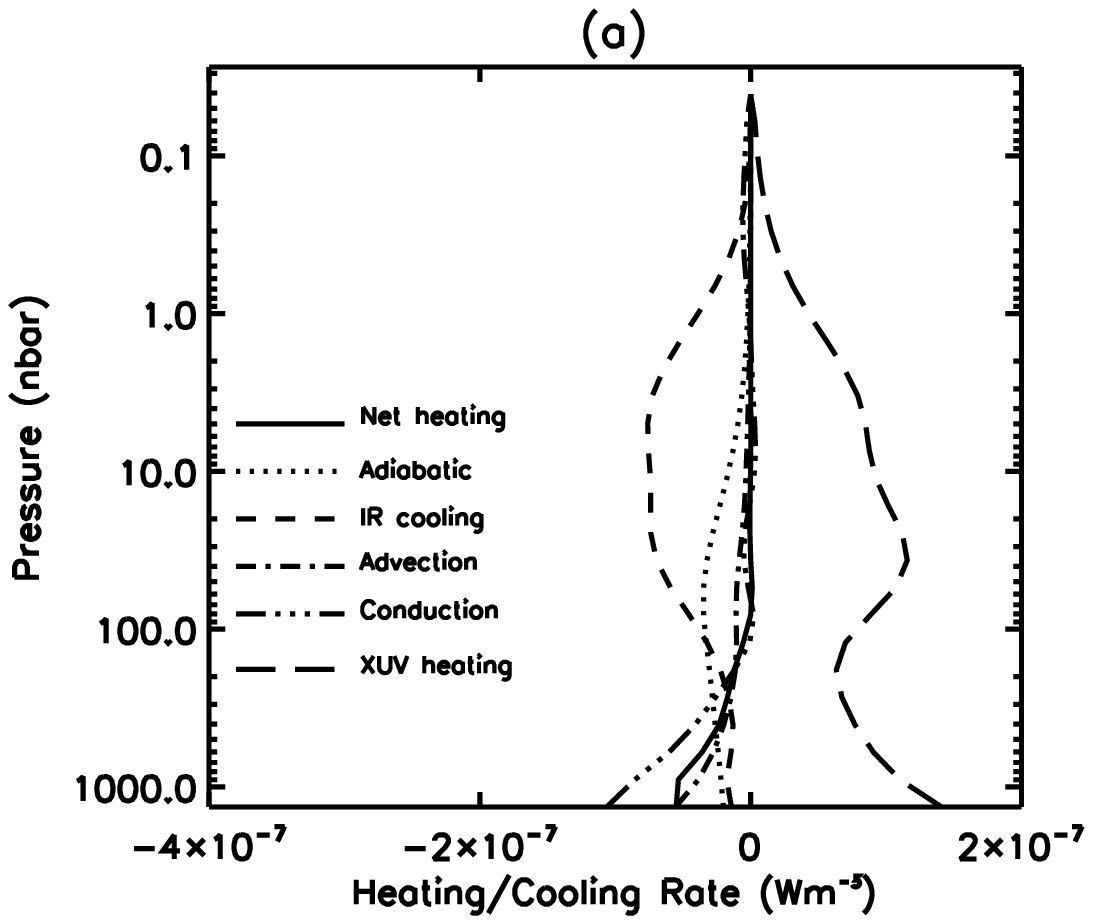}
\plotone{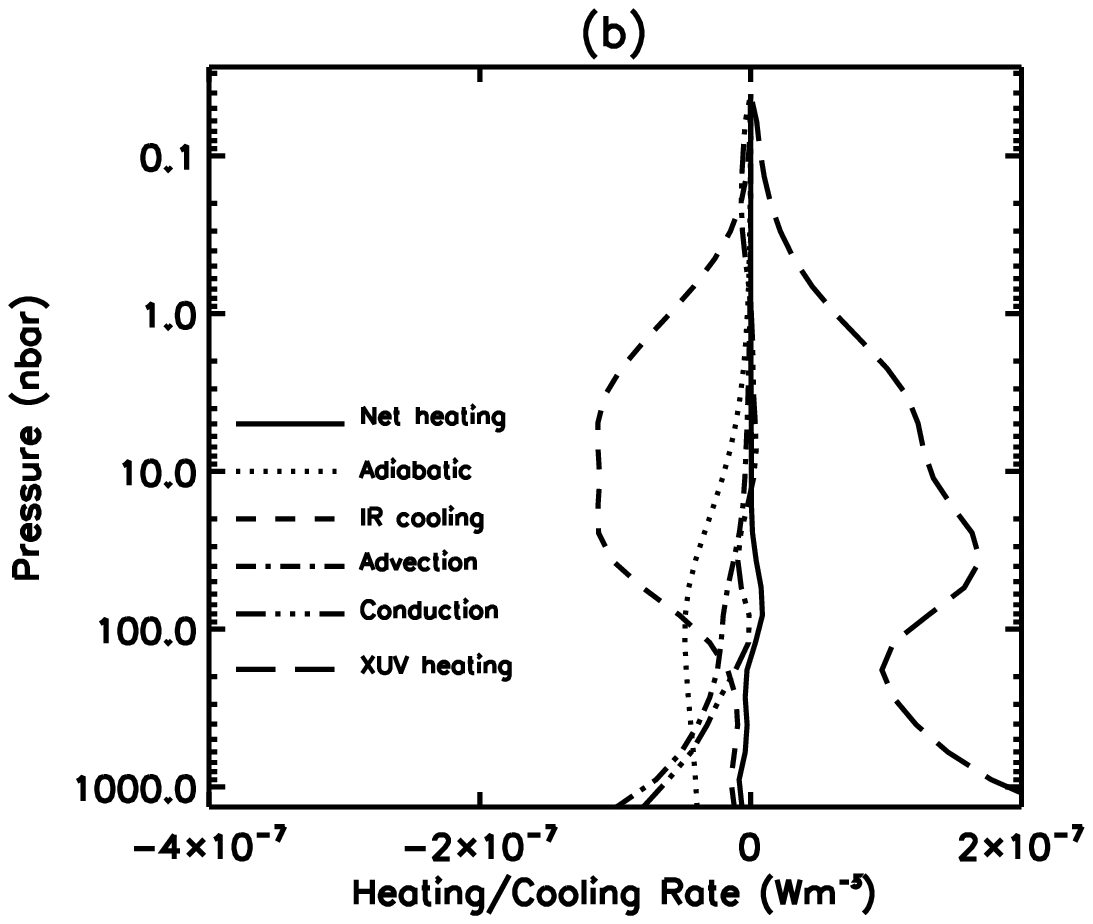}
\plotone{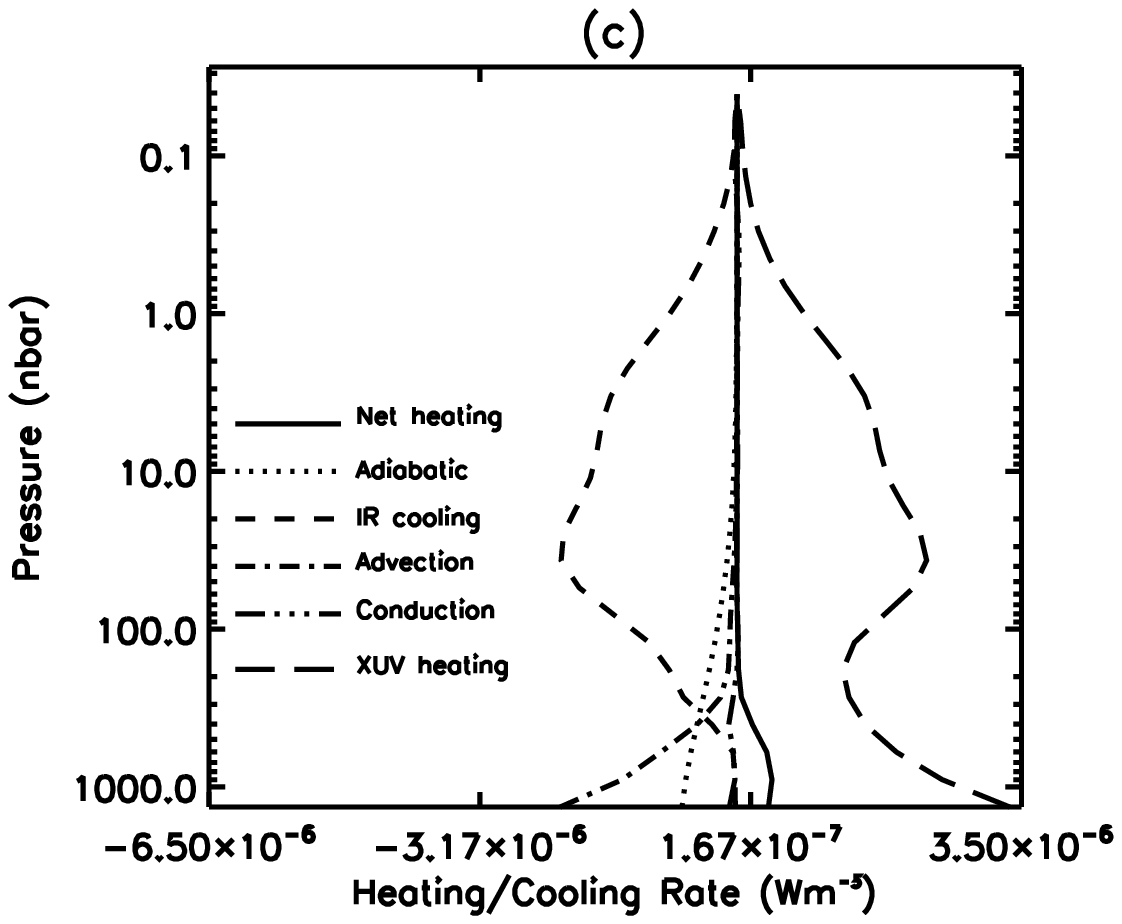}
\plotone{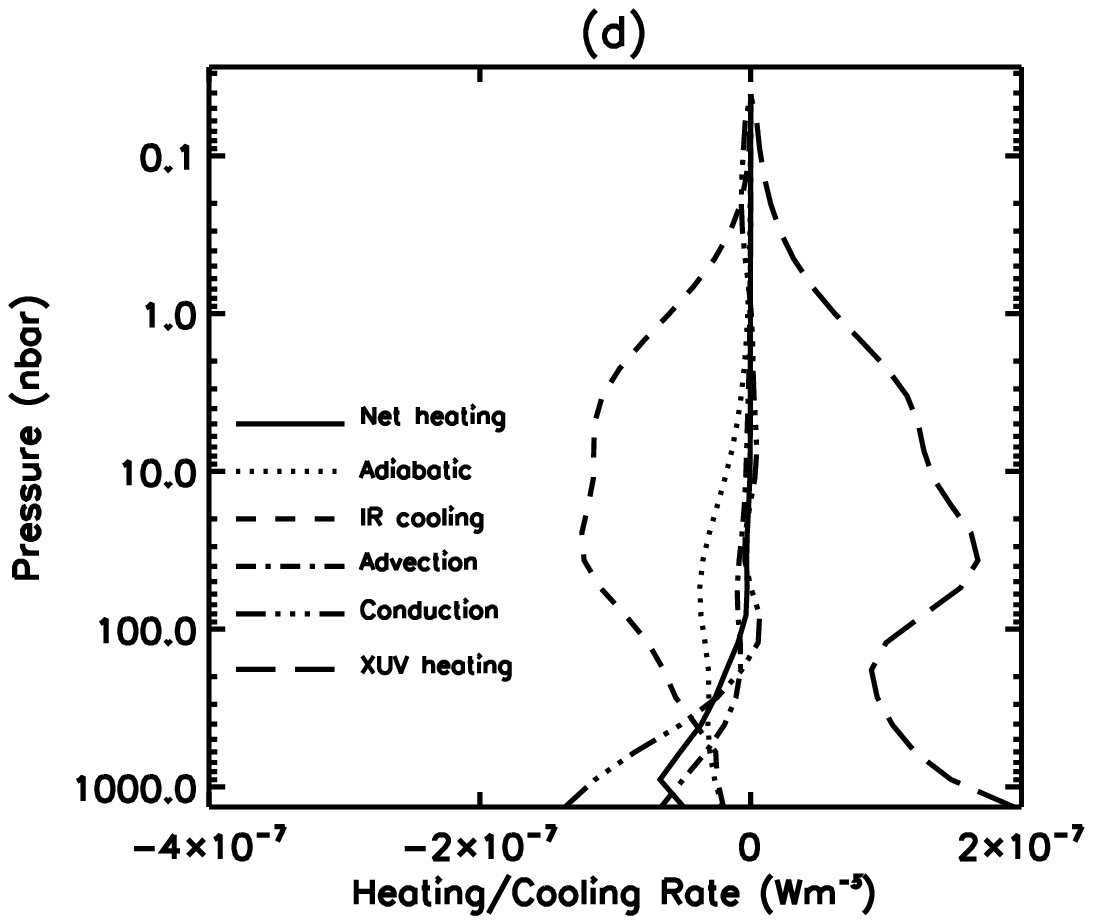}
\caption{Substellar energy equation terms (per unit volume) for the Exo-1 
simulations at (a) apastron, (b) $\theta =$~-153$^o$, (c) periastron, and 
(d) $\theta =$~153$^o$. \label{fg:ex01energy}}
\end{figure}

Figure~\ref{fg:ex01energy} shows the substellar heating and cooling rates by volume for  
the Exo-1 simulations at the same orbital positions.  Generally, in the middle and 
upper thermosphere (at $p <$~0.1 $\mu$bar), the XUV heating is effectively balanced 
by H$_{3}^{+}$ cooling.  However, at pressures lower than 1.0 nbar, this cooling 
is less important due to non-LTE conditions, and the heating is balanced mainly by 
conduction of heat downward.  In the lower thermosphere, the heating is balanced by 
the `triad' of vertical advection, adiabatic expansion, and conduction.  Due to 
relatively low temperature, H$_{3}^{+}$ cooling is not that significant near the 
lower boundary.  In the night side (not shown), the thermosphere is mainly heated 
by vertical downwelling and adiabatic contraction.  The heating is balanced by 
vertical conduction.    

In Figure~\ref{fg:ex01energy}, the solid line shows the net heating rate 
(the sum of all the energy equation terms) at each orbital position.  
The net heating rate is close to zero in the upper thermosphere, while variations 
take place in the lower thermosphere.  The radiative timescale in the thermosphere 
is relatively short, and in those regions where H$_{3}^{+}$ cooling is significant, 
any added heating is quickly balanced by an adjustment in the cooling rate.  This 
explains the rapid convergence of the Exo-1 simulations to steady state after 
only a few orbits.  Circulation in the upper thermosphere is driven by uneven 
stellar irradiation, which explains the relatively steep diurnal temperature 
gradient.  

In the lower thermosphere, on the other hand, net heating occurs during periastron.  
By the time the planet reaches $\theta =$~153$^o$ (0.22 AU), the lower 
thermosphere has started to cool down, due to conduction, advection and adiabatic 
expansion.  The cooling continues until the model reaches $\theta =$~-153$^o$.  
After this the model starts to heat up again as it moves towards periastron.     
The lower part of the model, instead of being radiation-driven, is dynamics-driven, 
and thus its response to the varying XUV fluxes is not as rapid as that of the 
upper thermosphere.  Also, the diurnal temperature differences tend to disappear 
at high pressures, and as we have seen, the circumplanetary wind appears.

We define the radiative cooling function as the ratio of the total infrared 
cooling rate to the total XUV heating rate.  This cooling function is relatively 
high for the Exo-1 simulations.  During apastron, it is 78 \%,  and the total 
H$_{3}^{+}$ infrared emission rate (in all spectral lines) is 6.1 $\times$10$^{14}$ W.  
At $\theta =$~-153$^o$, the cooling function is 72\% and the emission rate is
8.5 $\times$10$^{14}$ W.  During periastron, the corresponding figures are 
85 \% and 1.7 $\times$10$^{16}$ W, respectively.  Near the antitransit, and thus 
the possible secondary eclipse, the cooling function is 91 \% and the emission rate is 
1.1 $\times$10$^{15}$ W.  As a proxy for potentially detectable spectral lines,   
\citet{shkolnik06} estimated that the total H$_{3}^{+}$ emission rate 
should be of the order of 10$^{17}$-10$^{19}$ W if emissions are to be detected with 
current ground-based technology.  This estimate is based on a sample of F, G, K, and 
M stars within a distance of $\sim$40 pc.  
Unfortunately, even for these relatively high cooling 
functions, the predicted emission rates fall well short of the detection limits.      

\begin{figure}
\epsscale{1.0}
\plotone{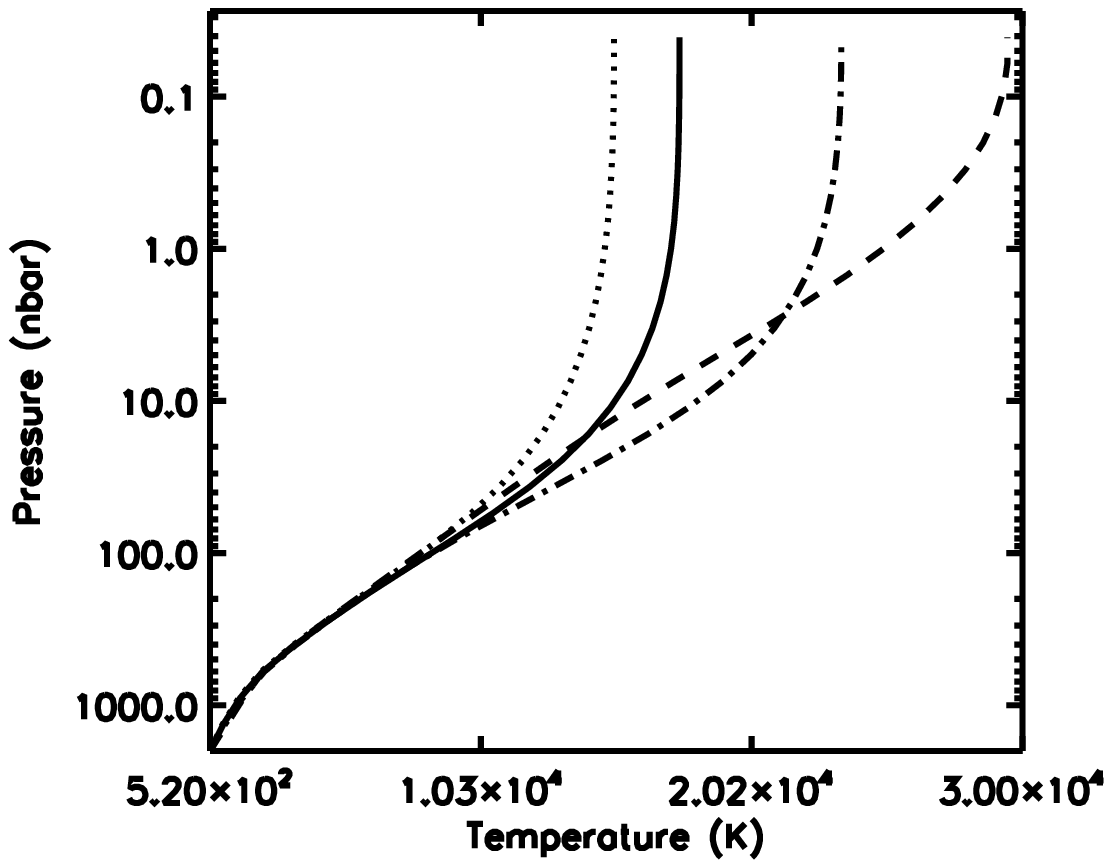}
\caption{Substellar P-T profiles for the Exo-2 simulations at apastron (solid line),
$\theta =$~-153$^o$ (dotted line), periastron (dashed line), and $\theta =$~153$^o$
(dash-dotted line). \label{fg:ex02pt}}
\end{figure}

Figure~\ref{fg:ex02pt} shows the substellar P-T profiles for the Exo-2 simulation at the 
same four orbital positions.  Outside the periastron, the P-T profiles are 
isothermal at pressures lower than about 0.7 nbar, and during periastron the profile 
is isothermal at pressures lower than about 0.1 nbar.  In the lower thermosphere, the 
temperature increases steadily with altitude.  The temperature in the upper thermosphere 
increases as the planet moves towards periastron from $\theta =$~-153$^o$.  The heating 
goes on for a while after periastron, but by the time the planet reaches $\theta =$~153$^o$, 
the outer layers have started to cool down.  Curiously, this does not apply to the 
region between 3.0 and 100 nbar, where the model is actually warmer at $\theta =$~153$^o$ 
than it is during periastron.  Towards apastron, the whole thermosphere cools 
down and this cooling continues until the planet reaches $\theta =$~-153$^o$ again.    

\begin{figure}
\epsscale{.45}
\plotone{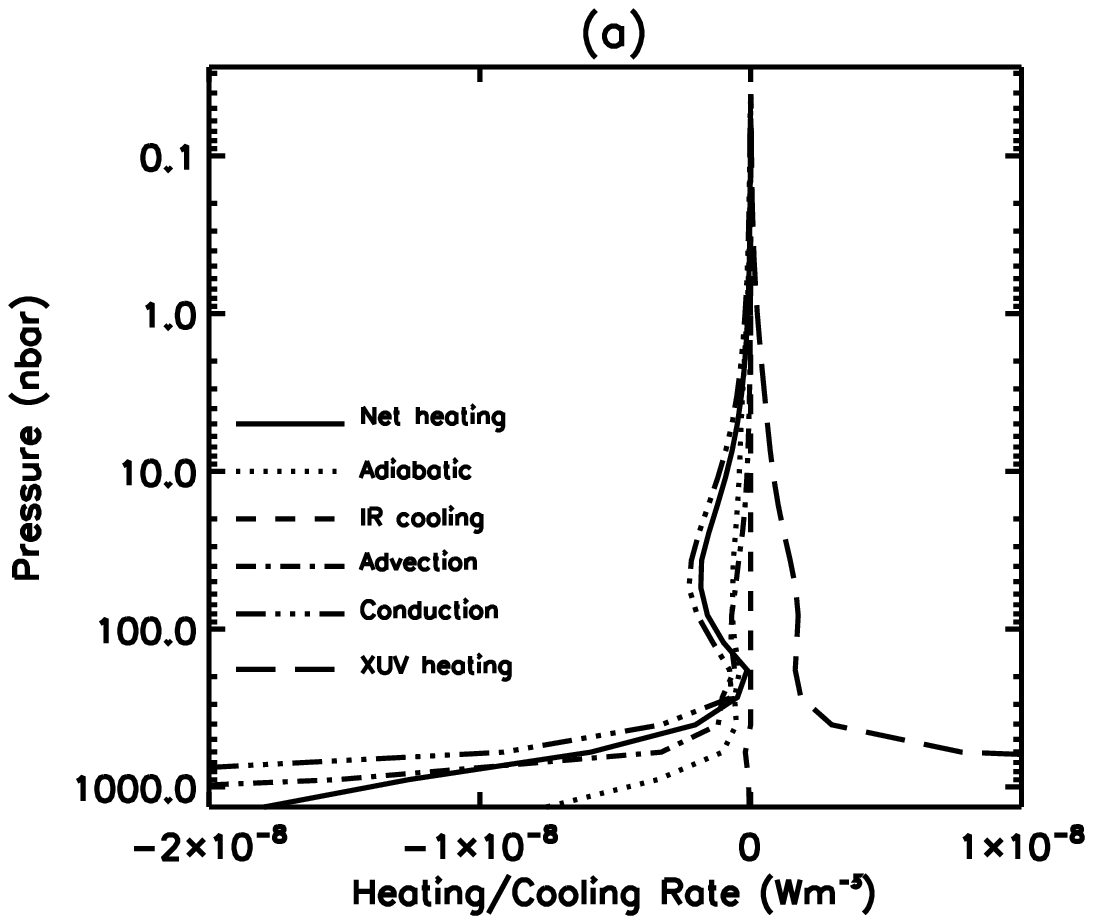}
\plotone{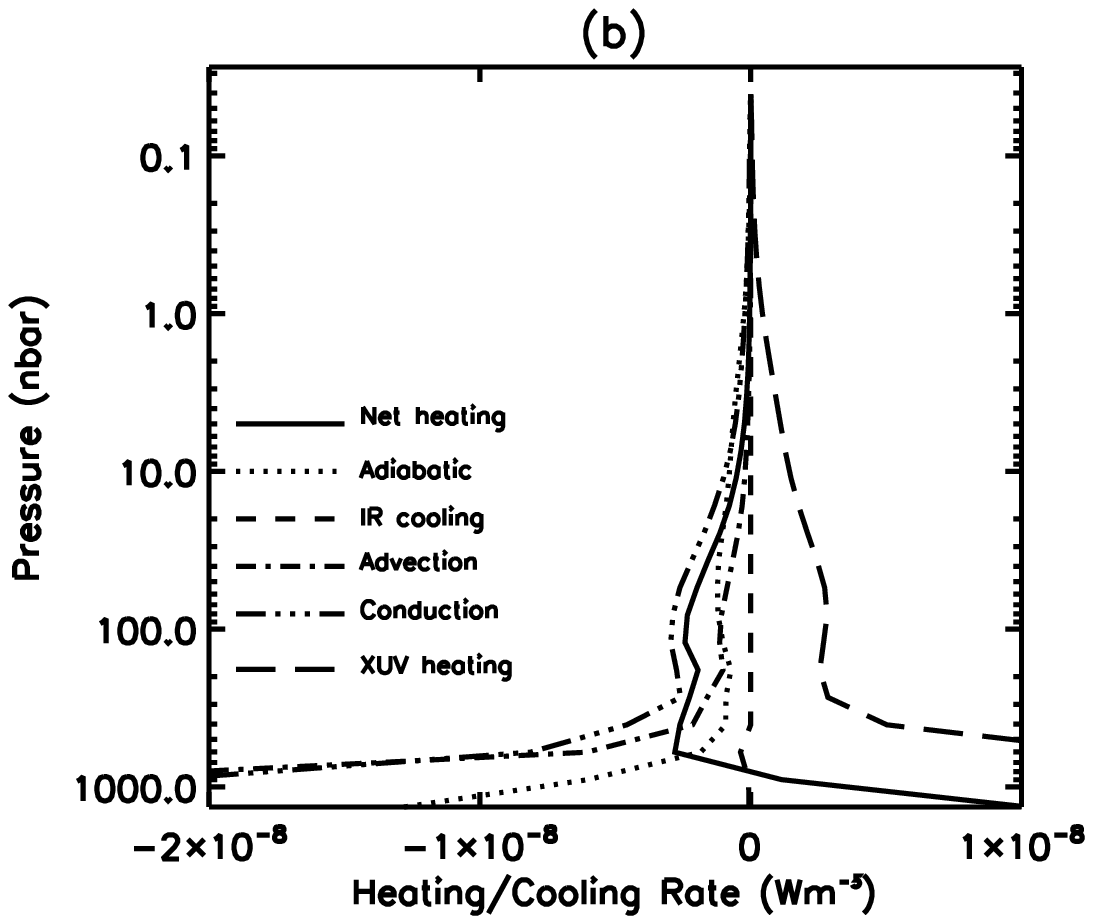}
\plotone{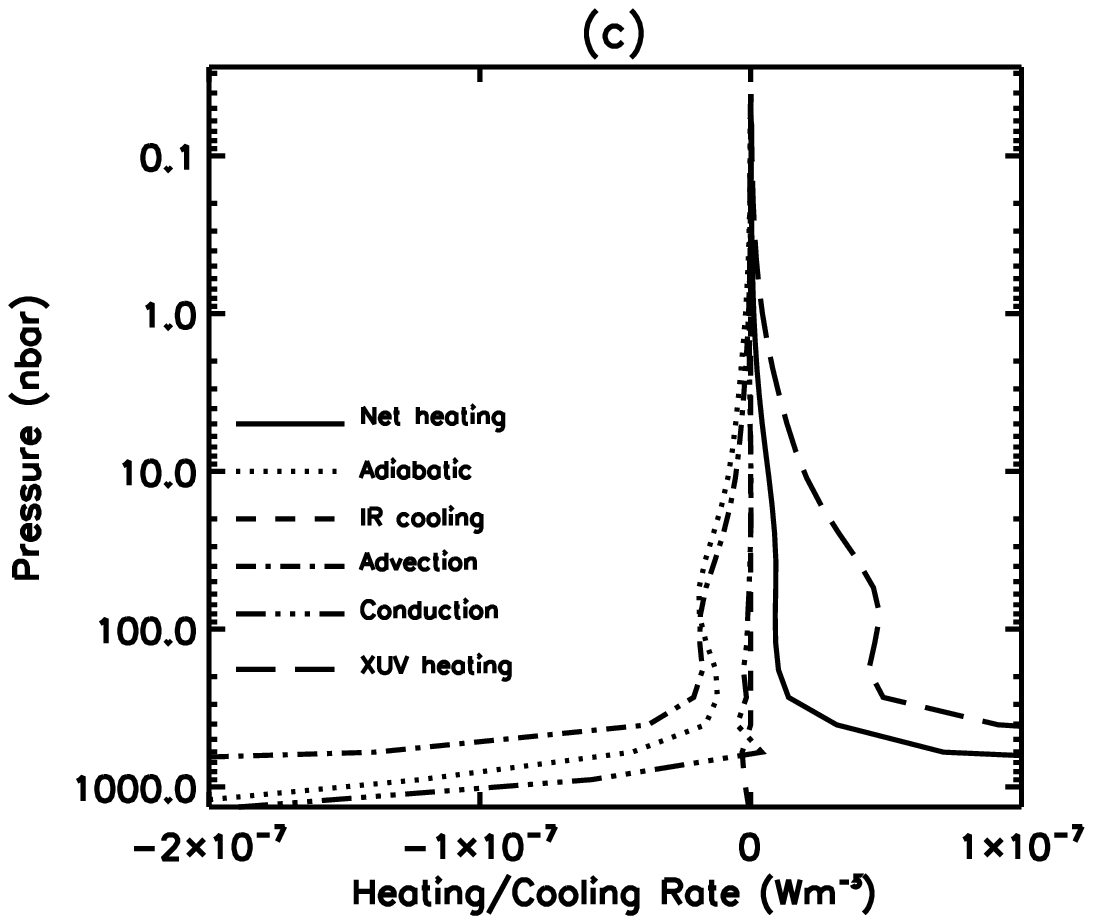}
\plotone{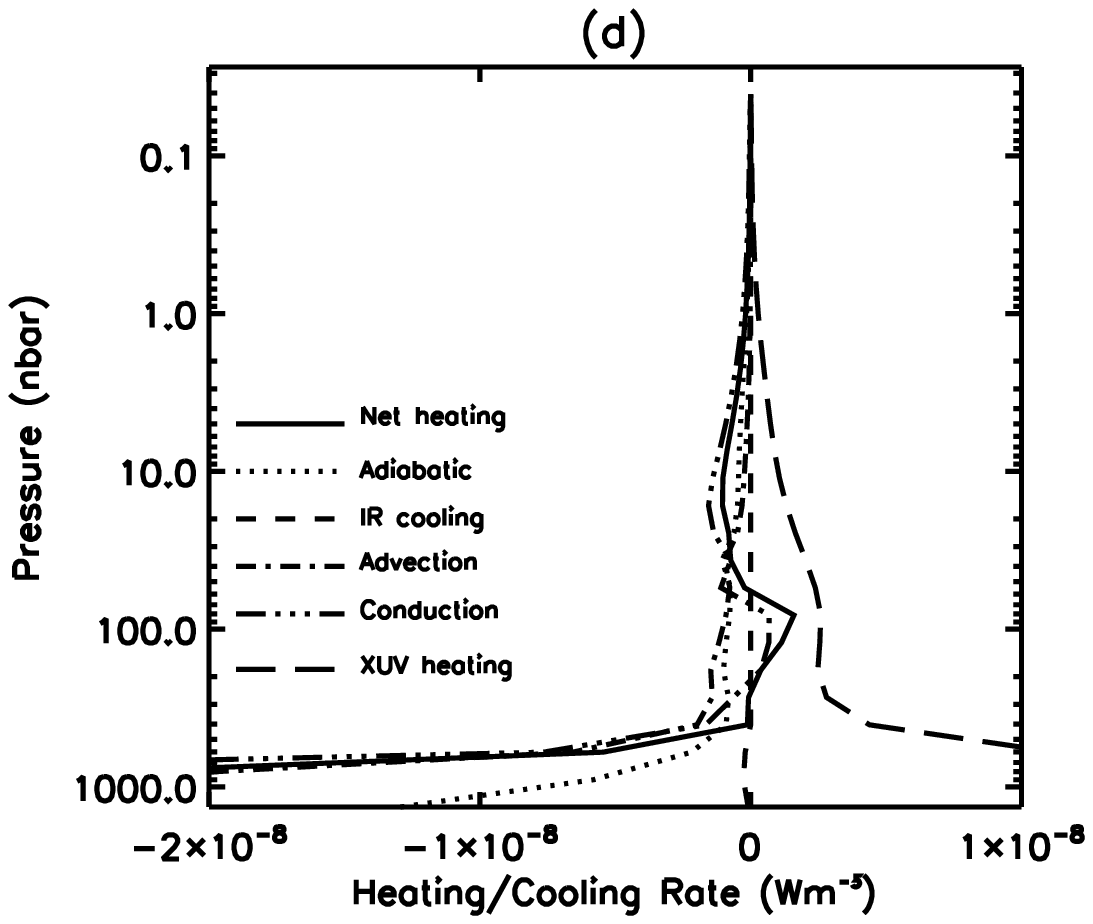}
\caption{Substellar energy equation terms (per unit volume) for the Exo-2 
simulations at (a) apastron, (b) $\theta =$~-153$^o$, (c) periastron, and 
(d) $\theta =$~153$^o$. \label{fg:ex02energy}}
\end{figure}

Figure~\ref{fg:ex02energy} shows the corresponding volume heating and cooling rates for  
the above P-T profiles.  The XUV heating (by volume) is concentrated near the bottom 
of the thermosphere.  This is because, compared to H$_2$, atomic hydrogen is not 
particularly effective in absorbing XUV radiation.  Also, the heating efficiency in the 
outer envelope is only 10 \%.  The radiation thus penetrates to the H$_2$-dominated 
bottom layers, where it is also absorbed by He, and where the heating efficiency is higher. 
However, the temperature is higher in the upper thermosphere, because the lower 
thermosphere is much denser and heats up sluggishly whereas even relatively inefficient 
heating is enough to produce high temperatures in the outer layers.         

Generally, the stellar heating is balanced by vertical conduction, vertical advection and 
adiabatic expansion.  In the night side, again, the heating is by downwelling and adiabatic 
contraction.  H$_{3}^{+}$ cooling is negligible compared to other cooling mechanisms in 
the dayside, and generally it only occurs at the bottom of the thermosphere.  
As indicated by the net heating rates, the thermosphere heats up as it approaches 
periastron after $\theta =$~-153$^o$, and the heating first occurs in the lower 
thermosphere.  The model reaches temperature maximum after periastron, and by the 
time the planet reaches $\theta =$~153$^o$, the lower and upper thermospheres have 
started cooling down.  Around the 100 nbar level XUV heating and vertical conduction 
are still heating the thermosphere at $\theta =$~153$^o$, and this is reflected in the 
P-T profiles above.    

The radiative cooling functions are low for the whole orbit.  At apastron, the 
cooling function is a meagre 0.23 \% and the total H$_{3}^{+}$ emission rate is  
4.9 $\times$ 10$^{11}$ W.  This is three orders of magnitude less than 
the emission rate for the Exo-1 apastron model.  The cooling functions for 
$\theta =$~-153$^o$, periastron, and $\theta =$~153$^o$ are 0.24, 0.097, and 0.19 \%, 
respectively.  The corresponding H$_{3}^{+}$ emission rates are 7.6 $\times$10$^{11}$ W, 
5.6 $\times$10$^{12}$ W, and 6.3 $\times$10$^{11}$ W.  

\subsection{Composition}
\label{subsc:composition}              

\begin{figure}
\epsscale{.70}
\plotone{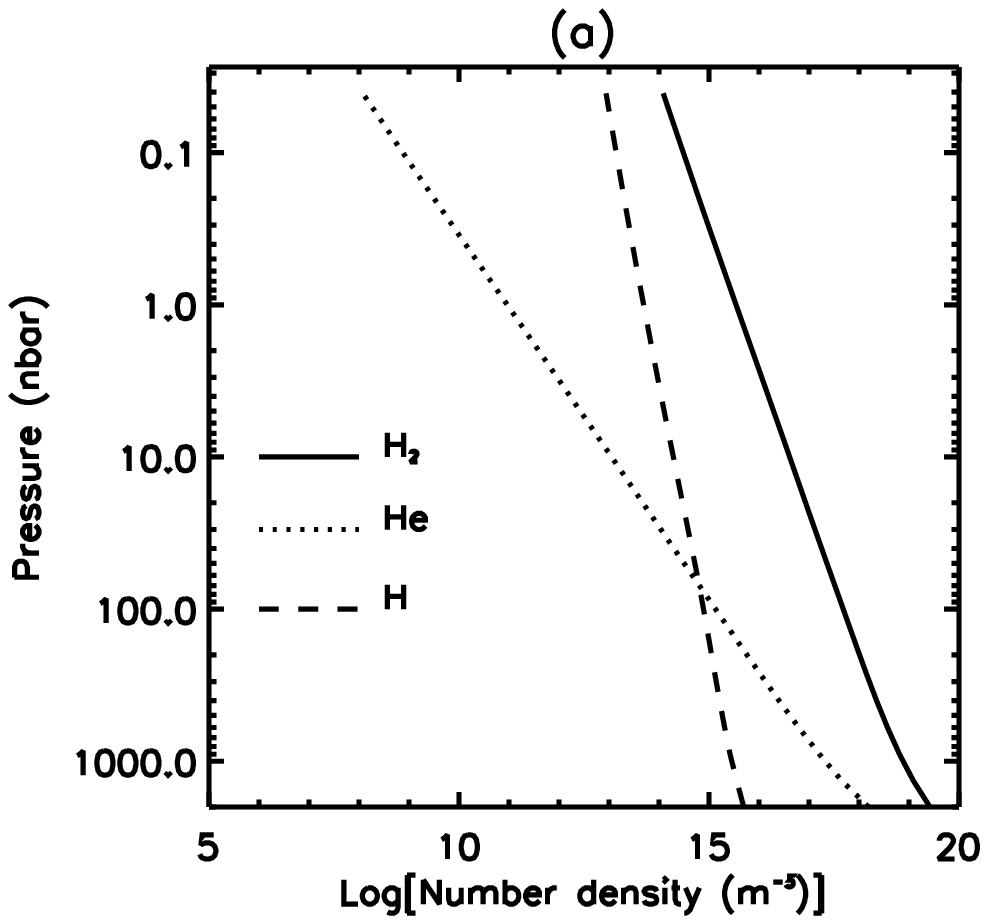}
\plotone{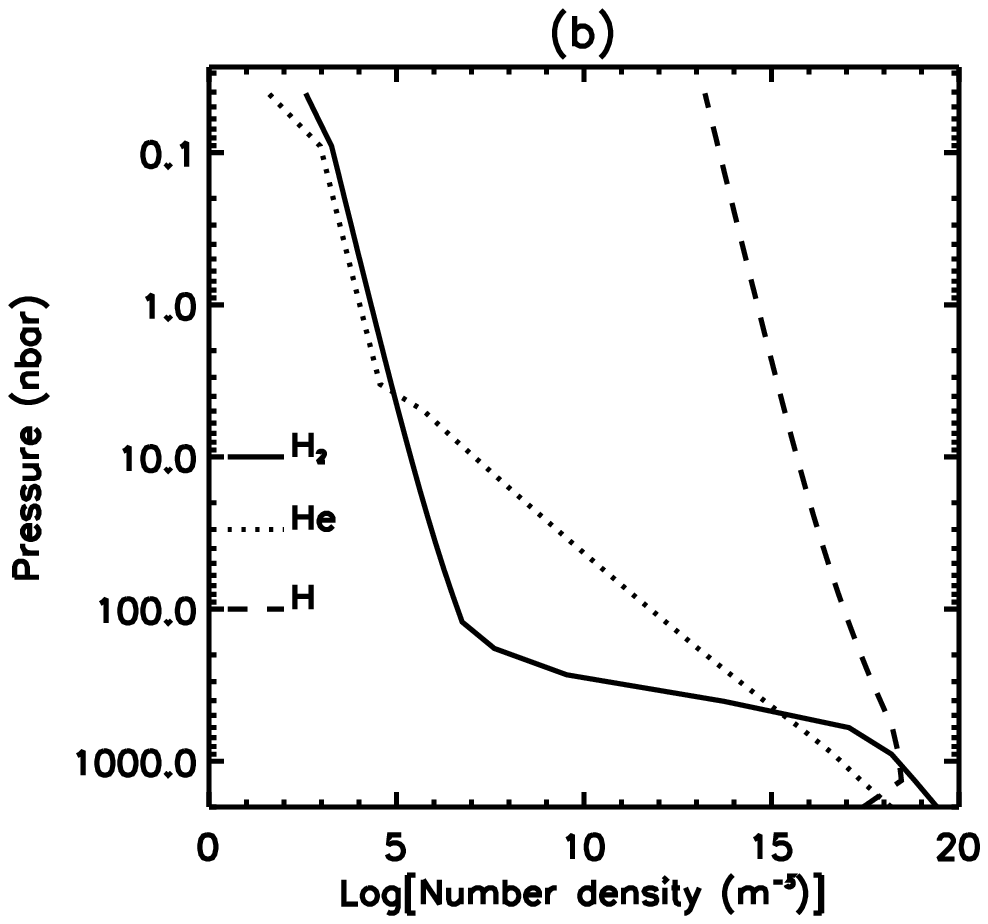}
\caption{Substellar neutral number densities during apastron for (a) Exo-1, 
and (b) Exo-2. \label{fg:ex0102compo}}
\end{figure}

Obviously, there are huge differences in thermospheric composition between the two 
scenarios, Exo-1 and Exo-2, with potentially observable consequences.  
Figure~\ref{fg:ex0102compo} shows the substellar number densities of the neutral 
species during apastron for both models.  For Exo-1, H$_2$ dominates at all levels.  
Atomic hydrogen overtakes helium around 0.1 $\mu$bar as the second most prominent 
neutral species.  Qualitatively, this picture prevails throughout the whole orbit.  

For Exo-2, atomic hydrogen is the dominant species at pressures lower than 1 $\mu$bar, 
and at higher altitudes the concentrations of the other species are negligible.  
Intriguingly, helium dominates over H$_2$ at pressures between $\sim$4 and 600 nbar.  
This is because the decline of H$_2$ in the upper thermosphere is not 
merely due to molecular diffusion, but it is significantly enhanced by thermal 
dissociation.

\begin{figure}
\epsscale{.70}
\plotone{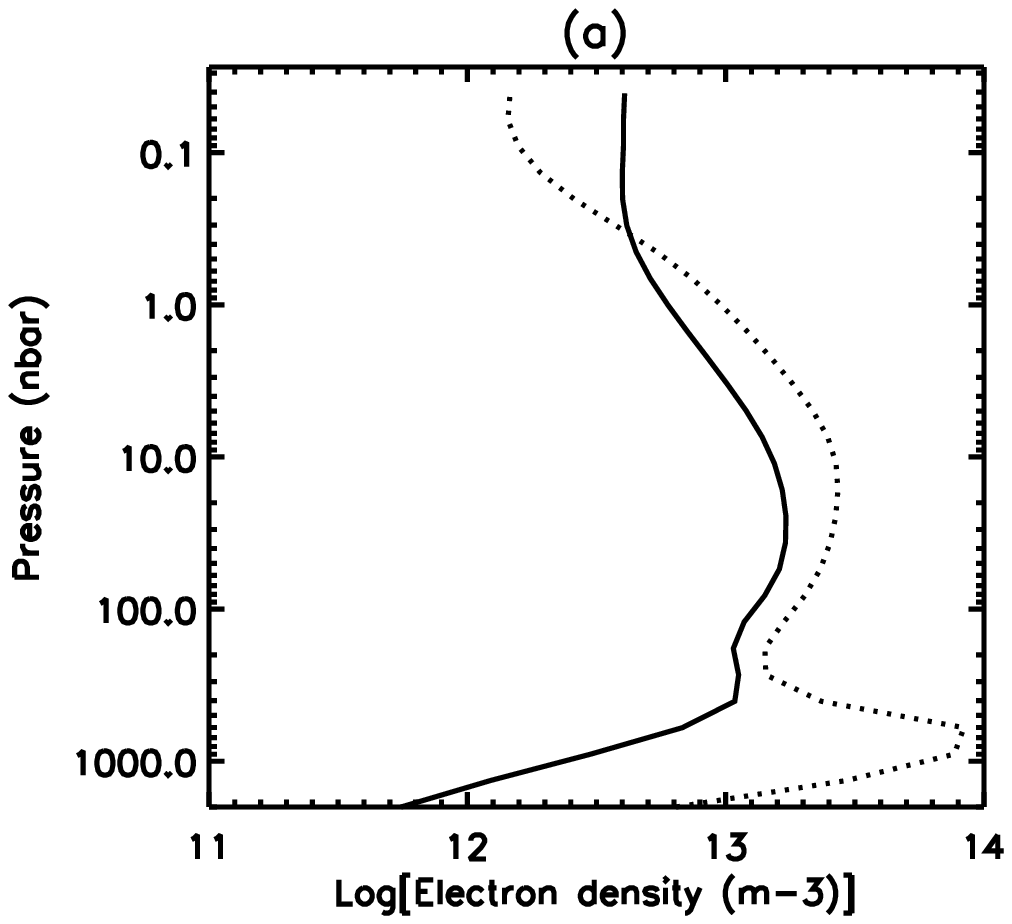}
\plotone{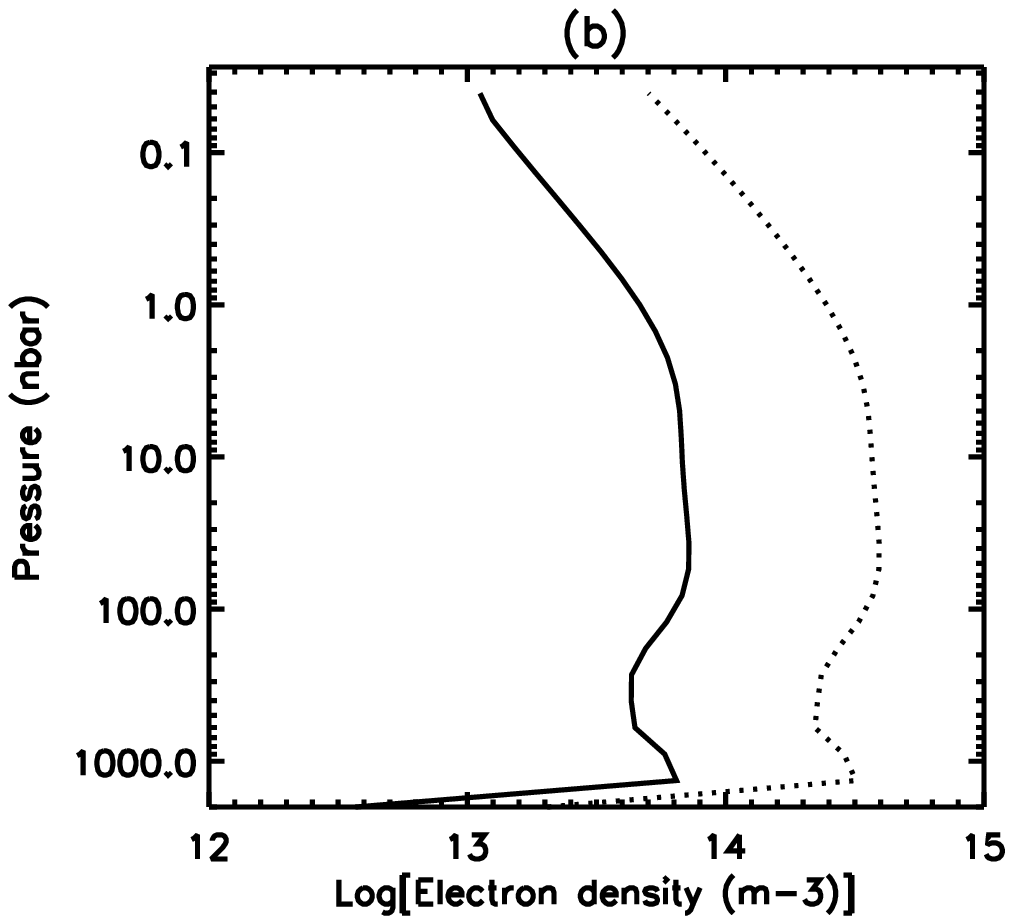}
\caption{Substellar electron density profiles for (a) Exo-1 and (b) Exo-2, 
during apastron (solid line) and periastron (dotted line). 
\label{fg:electron}}
\end{figure}

\begin{figure}
\epsscale{1.0}
\plottwo{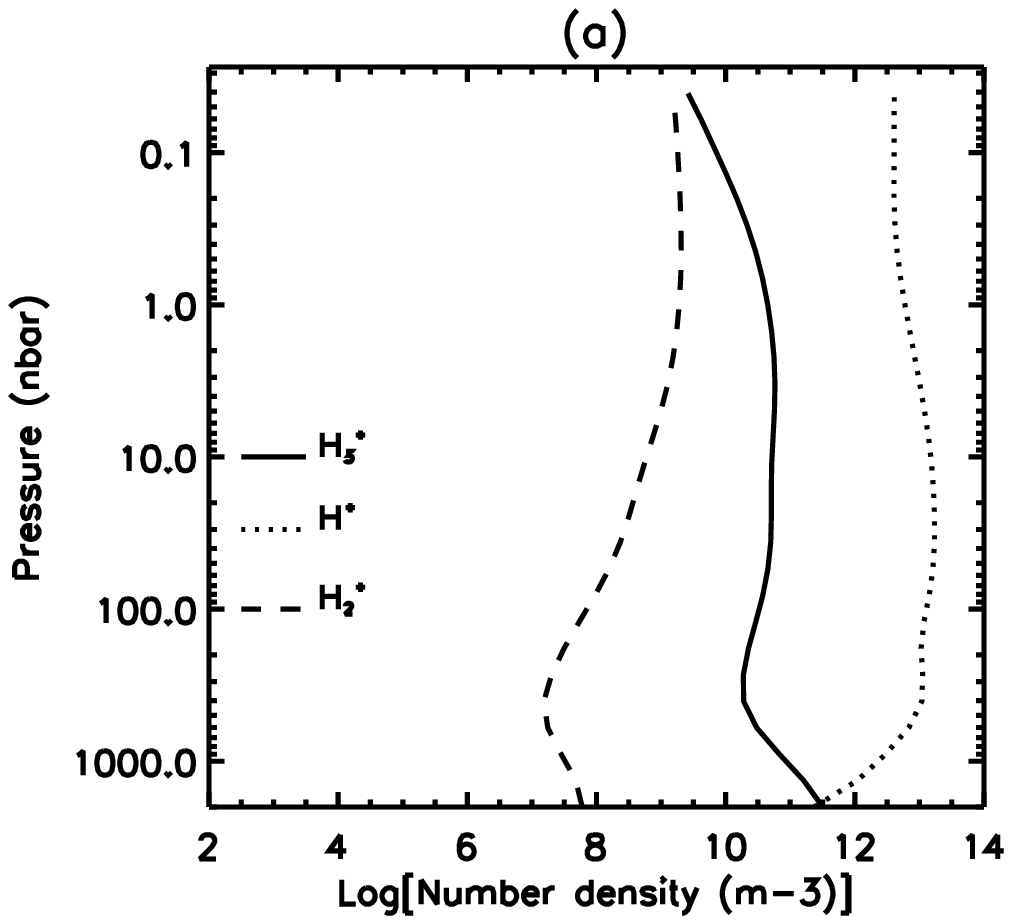}{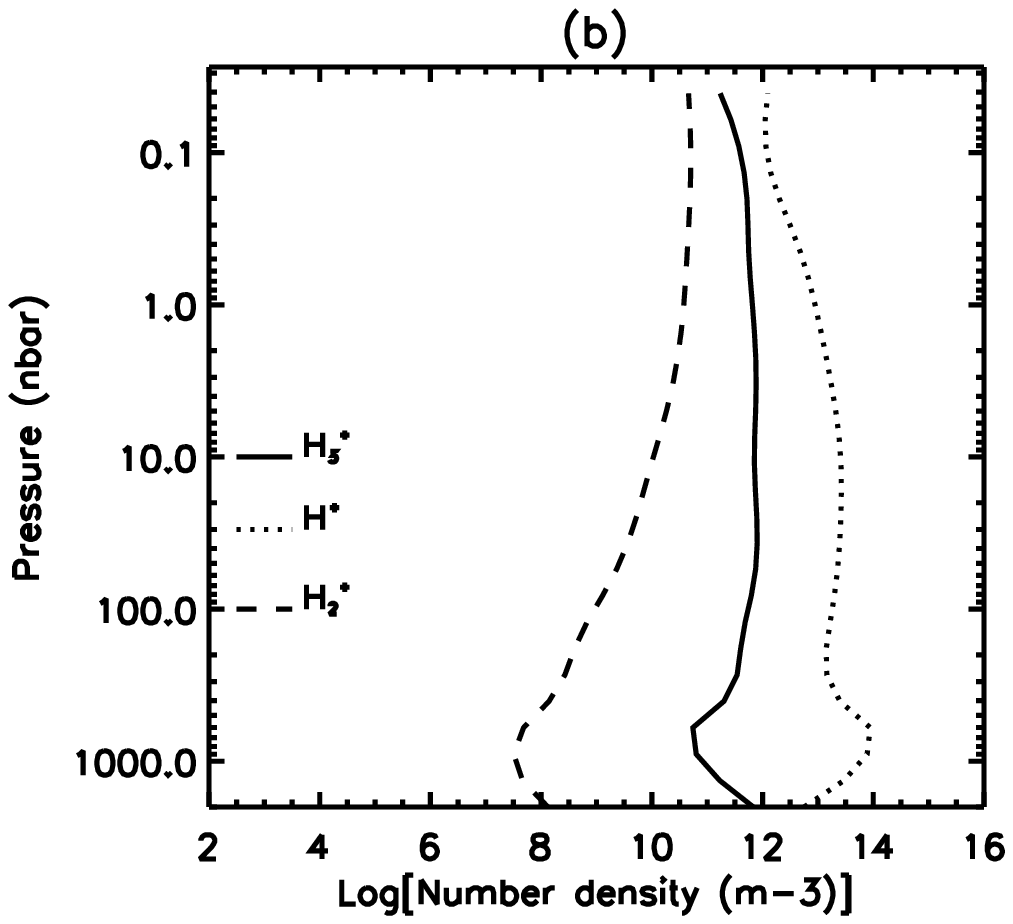}
\plottwo{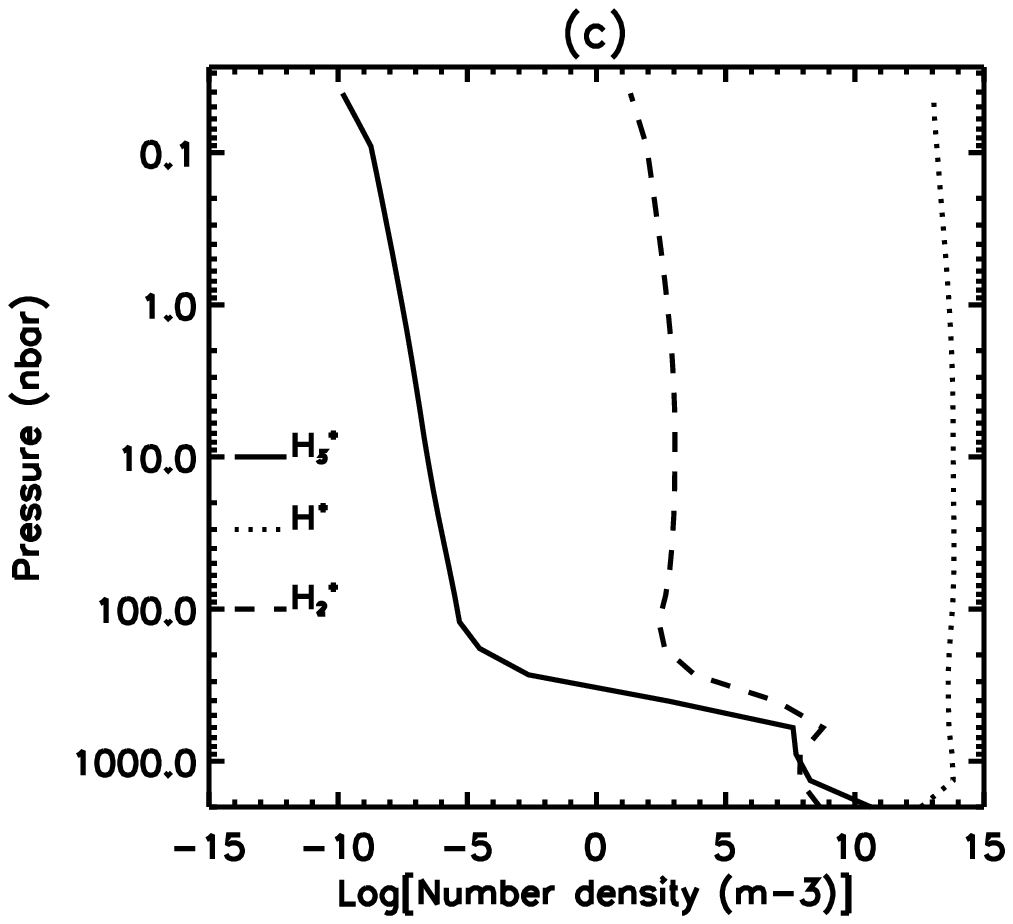}{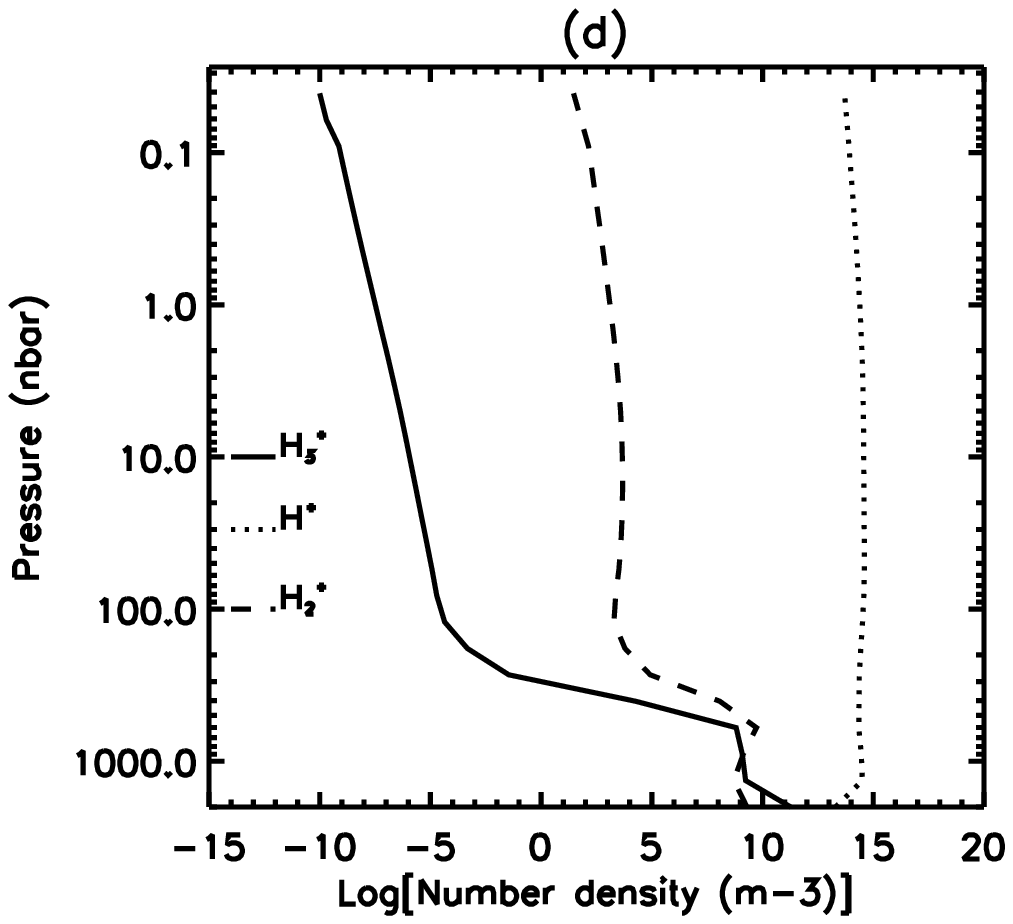}
\caption{Substellar ion density profiles for Exo-1 during (a) apastron and 
(b) periastron, and for Exo-2 also during (c) apastron and (d) periastron.
\label{fg:ions}}
\end{figure}

Figure~\ref{fg:electron} shows the substellar electron density profiles for Exo-1 
and Exo-2, both during periastron and during apastron, and Figure~\ref{fg:ions} 
shows the corresponding ions density profiles for H$^+$, H$_{2}^{+}$, and H$_{3}^{+}$.  
For Exo-1, the electron density during apastron increases with altitude in the lower 
thermosphere, and reaches a peak value of $\sim$1.8 $\times$10$^{13}$ m$^{-3}$ 
around 36 nbar.  In general, the electron density is close to the density of H$^+$, 
which is by far the dominant ion in the thermosphere.  The electron density drops 
steeply in the night side in the lower thermosphere, but in the outer layers, the 
densities are fairly uniform around the whole atmosphere.  This is because the 
lifetime of H$^+$ against recombination in the upper thermosphere is of the order 
of 40 hours, and the timescale for planetary rotation is similar.  The relatively 
long lifetime of H$^+$ indicates that the assumption of photochemical equilibrium 
is inaccurate in the upper thermosphere, where ion transport is likely to be 
important.  However, the lifetimes of other ions are relatively short, from seconds 
to minutes, and even the lifetime of H$^+$ is shorter in the lower thermosphere.  
The situation in this respect is thus similar to Jupiter, where H$^+$ densities 
also deviate from photochemical equilibrium in the upper thermosphere. 
The H$_{3}^{+}$ profile in Figure~\ref{fg:ions} has two distinct peaks, one 
between 0.7 and 5.0 nbar, and the other near the lower boundary.  
The substellar column densities of H$^+$ and H$_{3}^{+}$ 
are 7.9 $\times$10$^{18}$ m$^{-2}$ and 3.2 $\times$10$^{16}$ m$^{-2}$, 
respectively.

During periastron, there is a strong peak in the electron density profile 
near the 0.6 $\mu$bar level, with a density of 8.9 $\times$10$^{13}$ m$^{-3}$.  
This peak coincides with a H$^+$ peak in the lower thermosphere.  
Intriguingly, the electron and H$^+$ densities near the upper boundary are 
actually lower during periastron than they are during apastron.  
There are two main reasons for this.  First, the temperature in the upper 
boundary region is higher during periastron, and this leads to a lower 
overall density.  Second, the near-synchronisation of the planet's spin 
leads to enhanced day-night circulation that increases the mixing ratio of 
H$_2$ in the dayside.  The horizontal winds are fuelled by upwelling in the 
dayside and downwelling in the night side.  Upwelling brings up gas from 
the lower thermosphere, where the mixing ratio of H$_2$ is higher, and thus 
replenishes H$_2$ concentrations at higher altitudes.  Conversely, in the 
nightside the mixing ratio of atomic hydrogen is embellished as H$_2$ sinks 
with downwelling.  This kind of circulation is an important factor 
amplifying the cooling function in the thermospheres of tidally locked 
planets.  During apastron, the mixing of the atmosphere is more effective, as 
the relative rotation of the planet with respect to the star is faster. 
This leads to more uniform mixing ratios of atomic hydrogen, which do not 
exhibit notable diurnal variations.  For the periastron model, the substellar column 
densities of H$^+$ and H$_{3}^{+}$ are 2.0 $\times$10$^{19}$ m$^{-2}$ 
and 6.3 $\times$10$^{17}$ m$^{-2}$, respectively.  The H$_{3}^{+}$ 
densities are quite a bit higher than at apastron, and the mixing ratio 
of ion is also higher in middle and upper thermosphere. 
    
For Exo-2, during apastron, there are two peaks in the electron density 
profile at the bottom of the thermosphere and around 36 nbar, both with the 
roughly equal density of 6.3 $\times$10$^{13}$ m$^{-3}$.  Overall, the 
electron densities are higher than in the Exo-1 simulation.  As H$^+$ 
is virtually the only ion in the thermosphere, the electron density 
profile is effectively identical to the H$^+$ profile.  
The lifetime of H$^+$ in the upper thermosphere is around 90 hours, 
and a significant plasma density persists in the night side.  
This is such a large deviation from photochemical equilibrium that ion 
transport cannot be ignored.  Unfortunately this means that our density 
plots are not likely to be accurate.   
  
At pressures lower than about 0.6 $\mu$bar, there is virtually 
no H$_{3}^{+}$ in the Exo-2 simulation.  Very small amounts of H$_{2}^{+}$ are present, 
but the thermosphere does not contain enough H$_2$ to convert this into H$_{3}^{+}$.  In fact, 
the density of He$^+$ is generally higher in the thermosphere 
than the density of H$_{2}^{+}$.  The substellar column densities of H$^+$ and H$_{3}^{+}$ 
are 7.9 $\times$10$^{20}$ m$^{-2}$ and 5.0 $\times$10$^{14}$ m$^{-2}$, respectively.   

During periastron, the peak electron densities are much higher, around 
4.5 $\times$10$^{14}$ m$^{-3}$.  The relative ion concentrations are similar to 
the apastron model.  At pressures lower than $\sim$0.7 nbar, H$^+$ is actually the 
dominant species.  In fact, the mixing ratio of H near the upper boundary, in 
relation to H$^+$, is only 30 \%.  We have not made any provision for this in our simulations, which 
assume that the neutral density can be calculated from the ideal gas law, and 
that ionisation does not affect neutral mass fractions directly.  This is a 
further reason to treat the Exo-2 ion densities with suspicion, at least during 
periastron.  Elsewhere along the orbit, the density of H$^+$ tends to be much 
smaller compared to H.   

One last uncertainty related to the calculations of neutral and ion densities in the 
Exo-2 series is the fact that the numerical solution for the neutral mass 
fractions uses H$_2$ as the `dominant' species, although the equations themselves 
do \textit{not} assume a dominant species and diffusion is calculated for three 
major species instead of minor species diffusion.  However, the numerical solution 
evaluates the mass fractions 
for He and H, and then deduces those from unity in order to obtain the mass fraction 
for H$_2$.  This approach leads to numerical errors in the H-dominated part of the 
thermosphere.  The problem is limited to the relatively thin transition region between 
the H$_2$-dominated layers anf the H-dominated layers.
Above this region the mixing ratio of 
atomic hydrogen is unity for all practical purposes.  

\subsection{Variable temperatures}
\label{sc:varytemp}

\begin{figure}
\epsscale{0.7}
\plotone{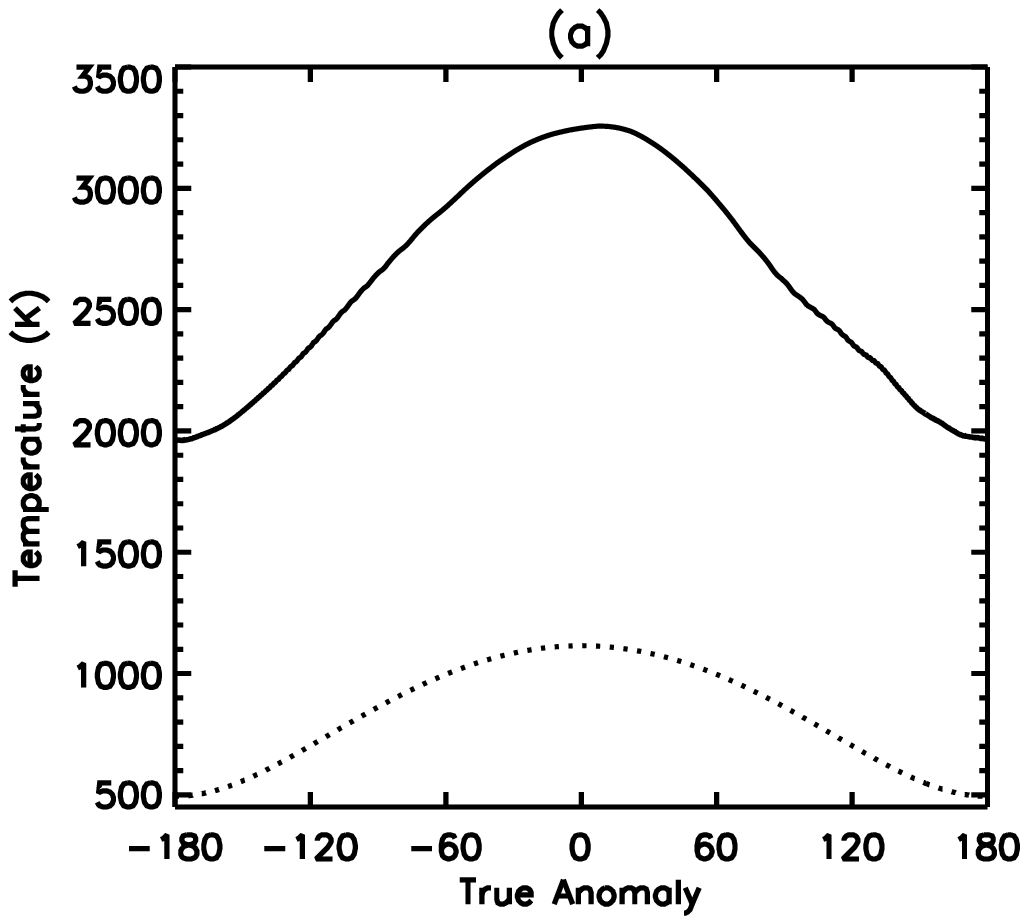}
\plotone{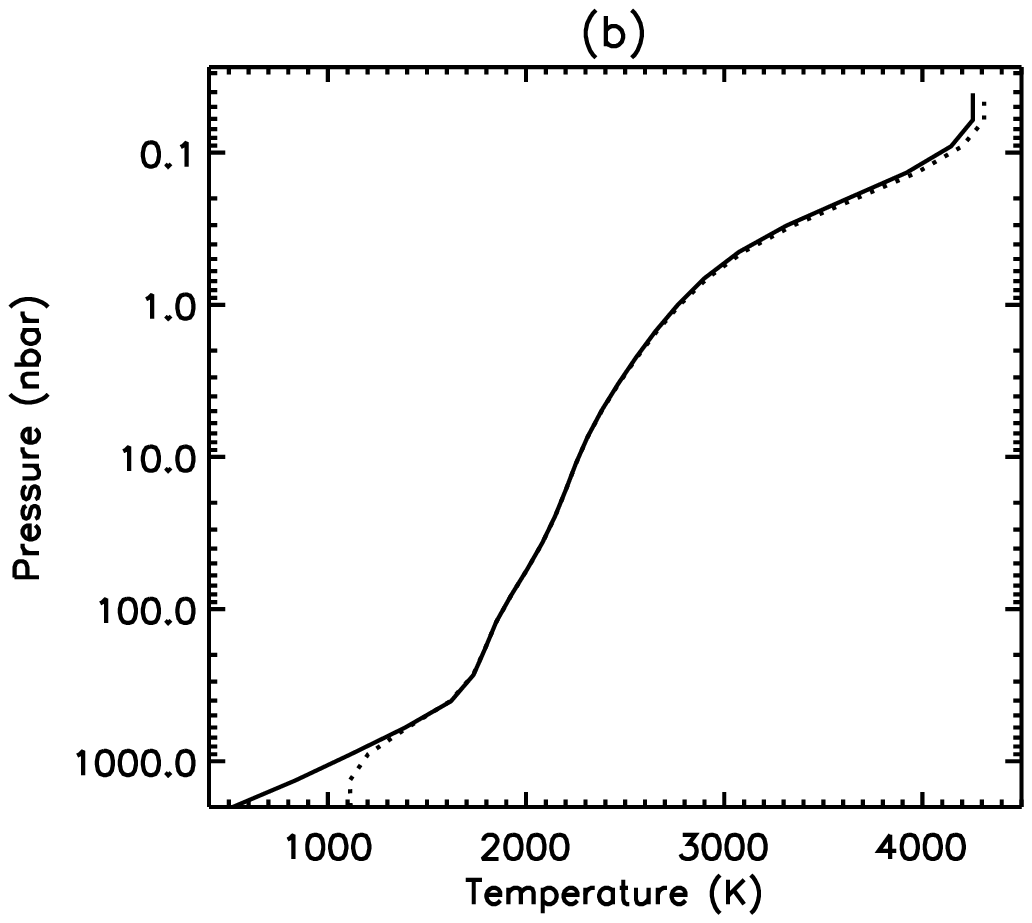}
\caption{(a) Globally averaged upper boundary (at 0.04 nbar) temperature 
(solid line) for the simulation with a variable lower boundary temperature 
(Exo-3), and the equilibrium temperature of a gas giant with a Bond albedo 
of 0.3 (dotted line). (b) Substellar P-T profiles for the Exo-1 (solid line) 
and Exo-3 (dotted line) simulations at periastron.
\label{fg:varytemp}}
\end{figure}

In order to explore the effect of variable lower boundary temperatures on our simulations, 
we generated a model, labelled Exo-3, which is otherwise identical to Exo-1 but where 
we varied the lower boundary temperature assuming that it is given by the equilibrium 
temperature of the planet (with a Bond albedo of 0.3) at different locations along the 
orbit.  We updated this temperature at every time step during the simulation.  
Figure~\ref{fg:varytemp} shows the lower boundary temperature as a function of true 
anomaly and contrasts this with the globally averaged temperature at the upper boundary 
of the Exo-3 simulation.  The equilibrium temperature varies from $\sim$490 K at apastron 
to just over 1,100 K at periastron.  This kind of variation, which is basically symmetric 
around the periastron, is unlikely to be realistic as the response of the lower atmosphere 
to varying stellar irradiation is likely to be more complex.

However, as Figure~\ref{fg:varytemp} indicates, varying the lower boundary temperature 
has almost no effect on the upper boundary temperatures.  As it turns out, changing the 
lower boundary temperature does not affect the qualitative nature of the circulation 
and composition of the Exo-1 simulation either.  Figure~\ref{fg:varytemp} also shows 
the substellar P-T profiles from the two simulations during periastron.  At pressures 
between about 0.6 $\mu$bar and 1 nbar the P-T profiles coincide.  Above the 1 nbar 
level, the Exo-3 model is slightly warmer, but the difference in temperature is barely 
noticeable.  In the lower thermosphere the temperatures are obviously different due to 
the different lower boundary conditions.  The XUV heating rates and the IR cooling 
rates in the two models are almost identical.  In the lower thermosphere the dayside 
cooling effect due to vertical advection is slightly reduced in the Exo-3 model, 
while the cooling effect due to adiabatic expansion is enhanced.  Also, due to the 
shallow vertical P-T profile near the lower boundary, the vertical conduction term 
produces slight heating near the lower boundary of the Exo-3 simulation.  In conclusion, 
it seems that the effect of the higher lower boundary temperature in the Exo-3 simulation 
is dwarfed by the fact that the temperatures in the thermosphere are largely determined 
by radiative balance, and most of the XUV radiation is absorbed above the lower boundary.

We did not run a separate simulation to explore how the varying lower booundary temperature 
would affect the Exo-2 simulations, but as the upper boundary temperature in those 
simulations is consistently above 10,000 K, it is unlikely that a variation of a few 
100 K at the lower boundary of the model would have a significant effect on the results.

\section{Conclusions and Discussion}
\label{sc:discussion}

We have generated three-dimensional simulations of the upper atmosphere of 
HD17156b, by using an improved version of the coupled thermosphere-ionosphere model for 
extrasolar giant planets (EXOTIM) \citep{koskinen07a,koskinen07b}.  By adopting the 
model for this EGP that orbits its G-type host star on a highly eccentric orbit 
($e =$0.67), we have been able to make predictions about the nature and stability of 
its upper atmosphere, thermospheric dynamics and ion concentrations.  These predictions 
have observable consequences and if followed up, they could prove very useful in probing 
the conditions in the atmosphere of this exotic world.

We have found that the composition of the thermosphere is sensitive to the composition 
of the underlying layers, and in particular to the mixing ratio of atomic hydrogen in 
the lower atmosphere.  Due to molecular diffusion, we do not anticipate a significant 
presence of heavy molecules such as CO, CH$_4$ and N$_2$, that would lead to extensive 
and complicated photochemistry \citep[eg.][]{garciamunoz07}, in the upper atmosphere.
The availability of H$_2$ regulates the H$_{3}^{+}$ cooling function in the thermosphere, 
and thus different mixing ratios of atomic hydrogen can lead to very different 
conditions in the upper atmosphere.  Unfortunately, the mixing ratio of atomic 
hydrogen in the lower atmosphere is poorly constrained.  Due to this uncertainty, 
we have generated two different types of HD17156b simulations: one with lower 
boundary mixing ratio of atomic hydrogen appropriate for Jupiter, and one with 
atomic hydrogen dominating the upper atmosphere.

In both cases, we find that the atmosphere of HD17156b remains stable everywhere 
along the orbit and does not begin to escape hydrodynamically.  This is despite 
the fact that a planet with a circular orbit at the periastron distance of HD17156b 
(0.052 AU) from the host star would almost certainly develop a planetary wind 
similar to that observed on HD209458b.  Evaporation is not as significant on 
HD17156b because the planet does not spend long enough within 0.1 AU from the 
parent star for hydrodynamic escape to begin.  Also, it is three times as massive 
as Jupiter, and thus the escape velocity and the thermal energy required for escaping 
the atmosphere is relatively high.    

Different thermospheric compositions, and thus different radiative cooling 
functions, lead very different conditions in the upper atmospheres.  If the thermosphere is 
dominated mostly by H$_2$, the H$_{3}^{+}$ cooling function varies between 70 and 
90 \% along the orbit, and the total infrared emissions are of the order of  
10$^{15}$-10$^{16}$ W.  In this case the extent of the atmosphere is negligible, and 
even the upper atmosphere stays relatively cool (2,000-3,000 K).  The thermal 
escape parameter at the upper boundary ranges from 400 to 600, and this renders 
any kind of evaporation negligible.  If, on the other hand, the thermosphere 
is dominated by H and H$^+$, the radiative cooling function is only 0.1-0.24 \%, 
and the total infrared emissions are of the order of 10$^{12}$-10$^{13}$ W.  
In this case the upper atmosphere is hot (15,000-30,000 K) and during periastron, 
the atmosphere expands beyond 1.4 R$_p$.  Depending on the orbital position, the 
thermal escape parameter ranges from 14 to 40, indicating that the atmosphere is 
in the Jeans escape regime.

These predictions can \textit{potentially} be verified by observations.  The infrared 
emissions from H$_{3}^{+}$ are potentially observable in the future, and they 
would provide the most direct clue to the properties of the upper atmosphere and 
its interaction with the host star.  Unfortunately, the predicted intensities are 
so low, that detection cannot be made with current ground-based technology 
\citep[eg.][]{shkolnik06}.  The extended atomic hydrogen cloud could possibly 
be observed by monitoring H Lyman $\alpha$ absorption during transit, which 
occurs near periastron.  It is not clear, however, if such an observation is 
possible in the near future because the instrument that performed a similar 
measurement on HD209458b, i.e. STIS onboard HST, is no longer available. 

Orbital dynamics and the planet's spin affect three-dimensional simulations in many 
important ways.  Circulation in the upper thermosphere is characterised by day-night 
winds, driven by a persistent diurnal temperature contrast.  This behaviour 
is enhanced during periastron, when the planet is in near-synchronisation with the star.  
In the lower thermosphere, temperatures are fairly uniform horizontally and the 
circulation is characterised by an eastward, circumplanetary jet.  In the H-dominated 
simulations, temperatures are generally more uniform and the wind speeds are slow.

As thermal escape is not as significant for HD17156b as it is for, say, HD2009458b, 
one wonders if non-thermal escape processes or tidal forces might come into 
play instead.  Non-thermal escape arises from charge exchange with the impinging stellar wind, 
photodissociation or dissociative recombination, acceleration of ions along open 
magnetic field lines, sputtering of neutral particles or ion pick-up by the stellar 
wind, and impact ionisation and dissociation by stellar wind particles.  
Quantifying these effects for a planet like HD17156b is difficult, because there 
is no information about its magnetic field or the properties of the stellar 
wind in its vicinity.  Estimates of non-thermal mass loss for a planet like HD209458b 
are of the order of 10$^8$-10$^9$ gs$^{-1}$ \citep{erkaev05}.  If similar values 
are to be expected for HD17156b, then evaporation from the planet is mostly non-thermal.     

Adopting the formula presented by \citet{erkaev07}, the approximate Roche lobe distance during 
periastron from HD17156b is 10-11 R$_p$.  This means that the occurence of so-called 
geometrical blow-off \citep{lecavelier04}, driven by the tidal forces between the planet and 
the star, is very unlikely.  In fact, it may be safe to ignore tidal forces in 
estimating the mass loss for HD17156b.   

\acknowledgments

T. T. K. has been supported by UCL and Perren studentships, while A. D. A. and S. M. have 
been supported by the UK Science and Technology Facilities Council (STFC). This work was partly 
carried out on the Keter High Performance Computer System, which is managed by the Miracle 
Astrophysics Project and funded by STFC.

\clearpage

\begin{deluxetable}{lll}
\tabletypesize{\scriptsize}
\tablecaption{Chemical reactions used by the model.\label{table:reactions}}
\tablewidth{0pt}
\tablehead{
\colhead{Reaction} & \colhead{Rate$^{a,b}$} & \colhead{Reference}
}
\startdata
1. H$_{2} + h\nu \rightarrow$ H$_{2}^{+} +$ e    &  -  &  \citet{yan98} \\
2. H$_{2} + h\nu \rightarrow$ H$^{+} + $H $+$ e  &  -  &  \citet{chung93},\citet{dujardin87} \\
3. H$_{2} + h\nu \rightarrow$ H$^+ + $H$^+ +$ e  &  -  &  \citet{dujardin87} \\
4. H $+ h\nu \rightarrow$ H$^{+} +$ e            &  -  &  \citet{hummer63} \\
5. He $+ h\nu \rightarrow$ He$^{+} +$ e          &  -  &  \citet{yan98} \\
6. H$_2 +$ M $\rightarrow$ 2H $+$ M  &  $1.5 \times 10^{-9} exp(-4.8e4/T)$  &  \citet{baulch92} \\
7. 2H $+$ M $\rightarrow$ H$_{2} +$ M  &  $8.0 \times 10^{-33}(300/T)^0.6$  &  \citet{ham70} \\
8. H$^{+} +$ H$_{2}(\nu \geq 4) \rightarrow$ H$_{2}^{+} +$ H  &  $1.0 \times 10^{-9} 
exp(-2.19e4/T)$  &  \citet{yelle04}, estimated \\
9. H$_{2}^{+} +$ H$_2 \rightarrow$ H$_{3}^{+} +$ H  &  $2.0 \times 10^{-9}$  &  \citet{thread74} \\
10. H$^{+} + $ H$_2 +$ M $\rightarrow$ H$_{3}^{+} +$ M  &  $3.2 \times 10^{-29}$  
&  \citet{kim94} \\
11. He$^{+} +$ H$_2 \rightarrow$ H$^+ +$ H $+$ He  &  $1.0 \times 10^{-9} exp(-5700/T)$  
&  \citet{moses00} \\
12. He$^{+} +$ H$_2 \rightarrow$ H$_{2}^{+} +$ He  &  $9.35 \times 10^{-15}$  &  \citet{anicich93} \\
13. H$_{3}^{+} +$ H $\rightarrow$ H$_{2}^{+} +$ H$_2$  &  $2.0 \times 10^{-9}$  &  
\citet{yelle04}, estimated \\
14. H$_{2}^{+} +$ H $\rightarrow$ H$^{+} +$ H$_2$  &  $6.4 \times 10^{-10}$  &  
\citet{kapras79} \\
15. H$^{+} +$e $\rightarrow$ H $+ h\nu$  &  $4.0 \times 10^{-12}(300/T_{e})^{0.64}$  &  
\citet{storey95} \\
16. H$_{2}^{+} +$e $\rightarrow$ H $+$ H  &  $2.3 \times 10^{-8}(300/T_{e})^{0.4}$  &   
\citet{auerbach77} \\
17. He$^+ +$e $\rightarrow$ He $+ h\nu$  &  $4.6 \times 10^{-12}(300/T_{e})^{0.64}$  &   
\citet{storey95} \\
18. H$_{3}^{+} +$e $\rightarrow$ H$_2 +$ H  &  $2.9 \times 10^{-8}(300/T_{e})^{0.64}$  &   
\citet{sundstrom94} \\
19. H$_{3}^{+} +$e $\rightarrow$ H $+$ H $+$ H  &  $8.6 \times 10^{-8}(300/T_{e})^{0.64}$  &      
\citet{datz95} \\
\enddata
\tablenotetext{a}{Photo-ionisation rates are calculated explicitly by using the  
photoionisation cross sections given in the references.}
\tablenotetext{b}{Two-body rates are given in cm$^{3}$s$^{-1}$ and three-body rates are 
given in cm$^{6}$s$^{-1}$.  The electron temperatures are assumed to be the same as neutral 
temperatures.}
\end{deluxetable}

\begin{deluxetable}{cccc}
\tabletypesize{\scriptsize}
\tablecaption{Non-LTE correction factors \label{table:nonlte}}
\tablewidth{0pt}
\tablehead{
\colhead{Pressure (nbar)} & \colhead{Temperature (K)} & \colhead{Detailed balance} &
\colhead{Experimental}
}
\startdata
2000  &  520   &  1.00   &  1.00   \\
1340  &  833   &  1.00   &  1.00   \\
899   &  1112  &  1.00   &  1.00   \\
602   &  1382  &  0.998  &  1.00   \\
404   &  1614  &  0.986  &  1.00   \\
271   &  1732  &  0.971  &  1.00   \\
181   &  1792  &  0.956  &  1.00   \\
121   &  1849  &  0.934  &  1.00   \\
82    &  1925  &  0.907  &  1.00   \\
55    &  2008  &  0.870  &  0.878  \\
36    &  2083  &  0.824  &  0.782  \\
25    &  2147  &  0.767  &  0.696  \\
16    &  2201  &  0.702  &  0.618  \\ 
11    &  2252  &  0.630  &  0.548  \\
7.4   &  2311  &  0.556  &  0.484  \\
5.0   &  2382  &  0.480  &  0.426  \\
3.3   &  2465  &  0.404  &  0.374  \\
2.2   &  2555  &  0.330  &  0.327  \\
1.5   &  2654  &  0.258  &  0.284  \\
1.0   &  2766  &  0.194  &  0.245  \\
0.67  &  2899  &  0.138  &  0.210  \\ 
0.45  &  3074  &  0.0916 &  0.179  \\
0.30  &  3314  &  0.0554 &  0.150  \\
0.20  &  3615  &  0.0303 &  0.124  \\
0.14  &  3919  &  0.0155 &  0.101  \\
0.09  &  4144  &  0.0082 &  0.0794 \\
0.06  &  4254  &  0.0050 &  0.0602 \\
0.04  &  4254  &  0.0036 &  0.0428 \\
\enddata
\end{deluxetable}

\begin{deluxetable}{lc}
\tabletypesize{\scriptsize} 
\tablecaption{Planetary and orbital parameters for HD17156b \label{table:parameters}}
\tablewidth{0pt}
\tablehead{
} 
\startdata
\textbf{Planetary parameters$^{a}$}   &          \\
M$_p$ (M$_{Jup}$)                     &  3.1   \\
R$_p$ (R$_{Jup}$)                     &  0.96   \\
                                      &          \\
\textbf{Orbital parameters$^{a}$}     &          \\
Eccentricity                          &  0.67  \\
Semi-major axis (AU)                  &  0.16  \\
Inclination ($^o$)                    &  85.4   \\
Period (days)                         &  21.2  \\
Longitude of periastron ($^o$)        &  121  \\
                                      &          \\
\textbf{Lower boundary conditions}    &          \\
T$_0$ (K)                             &  520.0   \\
p$_0$ (Pa)                            &  0.2     \\
Gravity (ms$^{-2}$)                   &  87.0    \\
Mixing ratio of H (Exo-1)             &  2 $\times$10$^{-4}$  \\
Mixing ratio of H (Exo-2)             &  0.01    \\
\enddata
\tablenotetext{a}{With the exception of the planetary radius, the orbital and 
planetary parameters were taken from \citet{gillon07}.  Note that \citet{gillon07} 
have recently published and estimated radius of 1.23$_{-0.2}^{+0.17}$ R$_{Jup}$ 
for the planet.  The value given by \citet{irwin08} is 1.01~$\pm$0.09 R$_{Jup}$.  
Thus, given the error ranges, these two estimates coincide.  We have used a value 
that agrees with the estimates given by \citet{irwin08}.}
\end{deluxetable}      

\clearpage

\clearpage  

\clearpage

\clearpage

\clearpage

\clearpage

\clearpage

\clearpage

\clearpage

\clearpage

\clearpage

\clearpage

\clearpage 

\clearpage

\clearpage

\clearpage

\clearpage

\end{document}